# Domain-Gated Latent Diffusion: Generative Inverse Design of HMX-Class Energetic Materials with First-Principles Validation

Yehudit Aperstein [1,*] and Alexander Apartsin [2]
[1] Department of Intelligent Systems, Afeka Tel-Aviv College of Engineering, Tel-Aviv, Israel
[2] School of Computer Science, Faculty of Sciences, Holon Institute of Technology (HIT), Holon, Israel
* Correspondence: apersteiny@afeka.ac.il

**Abstract:** Energetic materials power mining, demolition, propulsion and airbags, yet today's compounds were designed decades ago. A successor must combine high energy release, low sensitivity to accidental initiation and a practical synthesis route, found within an astronomically large molecular space. Generative models are the natural search tool, but their training data is mostly untrustworthy: of ~66,000 molecules with recorded properties, only ~3,000 were measured or computed from first principles. Models trained on all of them imitate the rough estimates and propose molecules that collapse under real physics. We introduce *Domain-Gated Latent Diffusion (DGLD)*, a diffusion model that treats data reliability as an explicit design parameter: labels are sorted into four trust tiers, and only trustworthy ones steer generation, while the unreliable majority still teaches the model what a plausible molecule looks like. Learned controls tune performance, safety and viability independently, and every proposal passes a four-stage screen ending in quantum-chemical (DFT) audit. DGLD proposes 10 molecules unknown to PubChem that survive it. The best, 3,4,5-trinitro-1,2-isoxazole, matches benchmark explosives HMX and PETN on calculated detonation performance, is unlike its training set, and has a four-step synthesis route. Trust gating is chemistry-independent, applying wherever abundant weak data surrounds a reliable core.

## 1. Introduction

Energetic-materials performance gains translate directly into reduced propellant mass for launch and mining applications, safer and more efficient automotive gas-generators, and lower-mass payloads across the field, yet the canonical anchors of the field (HMX and CL-20) were developed decades ago and few new HMX-class compounds have been disclosed in the decades since. The space of synthesisable CHNO small molecules with the right balance of crystal density, oxygen balance (OB), heat of formation (HOF), and detonation kinetics is vast, but traditional discovery explores it slowly. Computational methods offer acceleration, yet each existing family has a fundamental limitation: empirical formulas (Kamlet–Jacobs [1]) are regime-limited; discriminative surrogates [2][3] score candidates but do not propose them; generative language models trained on energetic corpora memorise training data; and standard guidance methods [4] fail silently when the generative trajectory is as short as molecular generation requires. Compounding these challenges, the labelled corpus is tier-stratified: of ~66 k labelled rows only ~3 k carry a trustworthy experimental or density functional theory (DFT) label on the target detonation properties, with the rest supplying empirical-formula or Uni-Mol-surrogate estimates of sharply lower reliability.

We introduce *Domain-Gated Latent Diffusion (DGLD)*, a reliability-aware generative framework for inverse molecular design from tier-stratified property data. At training time, a four-tier label-trust hierarchy gates which examples drive the conditional gradient (experimental and DFT labels) and which train only the unconditional prior (Kamlet–Jacobs and surrogate labels), preventing miscalibrated data from corrupting the generation signal. At sample time, a multi-task score model with six property and safety heads injects per-step steering into the diffusion trajectory; each head acts as an independent on/off switch without retraining the backbone. Decoded candidates pass through a four-stage validation funnel: a SMARTS chemistry gate, a Pareto re-ranker, semi-empirical GFN2-xTB triage, and full DFT audit (B3LYP/6-31G(d) + $\omega$B97X-D3BJ/def2-TZVP).

DGLD yields 10 unique DFT-confirmed novel leads (11 lead cards; L3 and L16 are the same compound); the headline compound, trinitro-1,2-isoxazole (L1), reaches $\rho_{\text{cal}} = 2.09$ g cm$^{-3}$ and $D_{\text{K-J,cal}} = 8.25$ km s$^{-1}$, placing it within the HMX/PETN performance band at max-Tanimoto 0.27 to our 65 980-row labelled corpus (Section 2.2). DGLD behaves as a *productive-quadrant generator*: it samples molecules that are simultaneously novel relative to the labelled master and competitive with the HMX/PETN reference class on predicted detonation performance. Across three strong energetics-domain baselines (SMILES-LSTM, SELFIES-GA, REINVENT 4), DGLD is the only method whose novel candidates remain on-target under first-principles DFT validation: SMILES-LSTM memorises 18.3% of outputs exactly; SELFIES-GA returns 75% (75/100) corpus rediscoveries and its best novel candidate collapses from $D_{\text{surrogate}} = 9.73$ to $D_{\text{DFT}} = 6.28$ km s$^{-1}$ under DFT audit (a 3.5 km s$^{-1}$ surrogate artefact); REINVENT 4 generates novel high-N heterocycles but peaks at a surrogate $D = 9.02$ km s$^{-1}$ that was not DFT-audited. Separately, the unguided sampler independently recovered E1 (4-nitro-1,2,3,5-oxatriazole), a known oxatriazole (PubChem CID 57438703) rather than a novel structure, reaching $D_{\text{K-J,cal}} = 9.00$ km s$^{-1}$ from a chemically distinct scaffold family, pending thermal stability confirmation and an oxatriazole-class DFT anchor (Section 3). The tier-gating recipe is domain-agnostic; only the validation funnel changes per application.

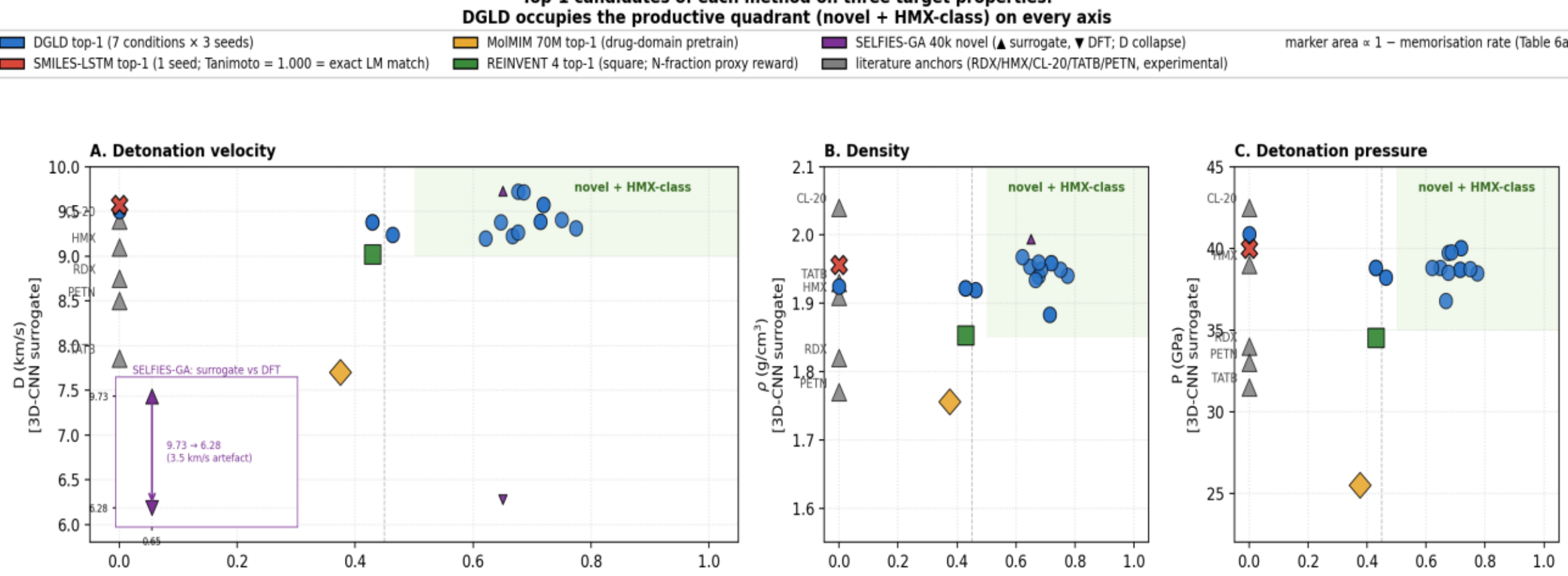

**Figure 1.** Top-1 candidate per method against novelty on three property axes ($D$, $\rho$, $P$). DGLD (blue, 7 settings × 3 seeds) clears the novelty floor (max-Tanimoto < 0.55) on every axis and lands in the HMX-class band (DGLD points on the $D$ panel are Uni-Mol surrogate scores and the anchors are at their reported literature $D$; only the DFT-calibrated Kamlet–Jacobs values for the DGLD leads, in Supplementary Table S30, are directly scale-matched to the anchors). SMILES-LSTM (red X) is exact rediscovery (Tanimoto = 1.0); MolMIM 70 M (gold) is novel but at $D = 7.70$ km s$^{-1}$; REINVENT 4 (green square) reaches $D = 9.02$ km s$^{-1}$ at novelty 0.43 (Table 2); SELFIES-GA (purple) shows a surrogate-to-DFT collapse on the $D$ panel inset ( $D = 9.73 \rightarrow 6.28$ km s$^{-1}$). Marker area encodes (1 − memorisation rate) , shrinking SMILES-LSTM and SELFIES-GA. Literature anchors (RDX/HMX/CL-20/TATB/PETN) at novelty = 0 (grey triangles); green-tinted upper-right quadrant is the productive zone.

*Note on Figure 1.* Hz-C2 top-1 surrogate values (Uni-Mol scale): $D$ = 9.39 km s$^{-1}$, $P$ = 38.7 GPa, within 5% of the strict targets. On the 6-anchor-calibrated K-J scale L1 reaches $D_{\text{K-J,cal}}$ = 8.25 km s$^{-1}$ and $P_{\text{K-J,cal}}$ = 32.9 GPa, slightly below the strict thresholds but in the HMX-class band by anchor-relative ranking.

1. *A tier-gated training recipe for sparse-label molecular generation.* Only trustworthy labels enter the conditional gradient, while lower-confidence labels train the unconditional prior (Section 4.1).
2. *A multi-task score model with selectable sample-time steering.* Six heads share a single trunk trained jointly over viability, sensitivity, hazard, performance, and two synthesis-complexity signals. At sample time, three heads supply per-step gradient steering on the latent; per-head scales act as on/off switches over chemistry classes without retraining the diffusion backbone.
3. *Self-distillation refinement of the viability head.* The viability head is refined by mining the model's own false-positive outputs as hard negatives against a chemist-defined held-out probe; the production budget-918 checkpoint is round 2 of self-distillation (Section 4.11), closing the gap between the Random Forest teacher's class boundary and the latent regions the sampler actually inhabits.
4. *Pool-fusion sampling for scaffold diversity.* Independent end-to-end sampling lanes (differentiated by denoiser checkpoint, conditioning targets, and guidance configuration) are unioned and deduplicated post-decode, making each axis an orthogonal diversity lever without increasing per-lane compute.

5. *DFT-validated novel leads in the HMX/PETN band (Kamlet–Jacobs relative ranking).* Eleven chem-pass leads are confirmed as true local minima under B3LYP/6-31G(d) and calibrated against six reference compounds under $\omega$B97X-D3BJ/def2-TZVP. Lead L1 (trinitro-1,2-isoxazole) is absent from PubChem, the SureChEMBL patent corpus [5], and the 65 980-row labelled master (max-Tanimoto 0.27); DFT-calibrated: $\rho_{\mathrm{cal}} = 2.09\ \mathrm{g\ cm^{-3}}$, $D_{\mathrm{K\text{-}J,cal}} = 8.25\ \mathrm{km\ s^{-1}}$; Uni-Mol surrogate: $D = 9.56\ \mathrm{km\ s^{-1}}$ (see Section 2.3 for K-J relative-ranking basis). The unguided sampler additionally recovered the known oxatriazole E1 (above); the extension set spans eight Bemis–Murcko scaffolds across seven chemotype families, of which E1, E6, and E9 are known compounds.

Taken together, these contributions sit on three axes of generative modelling for chemistry discovery: how a diffusion prior should consume unevenly reliable property labels, how the generative loop couples to quantum-chemical simulation as a validation stage rather than a post-hoc check, and what either buys on an advanced-materials target whose design objectives (energy release, stability, and synthesisability) are in direct conflict.

### 1.1. Related Work

DGLD sits at the intersection of molecular generative modelling, diffusion models with classifier-style guidance, and property prediction for energetic materials; an extended survey is given in the Supplementary Note (Extended Related Work). VAE-based generators learn a continuous latent over string or graph representations and navigate it for property optimisation [6][7]; we adopt the syntactically robust SELFIES representation [8] and fine-tune the LIMO MLP-VAE [9] as the frozen encoder; MolMIM [10] and the RL-based REINVENT 4 [11][12] are among the comparative no-diffusion baselines (Section 4.16). On the graph side, DiGress [13] performs discrete denoising diffusion directly on molecular graphs, achieving exact validity at the cost of re-introducing property conditioning through a separate guidance term. On the generative-prior side, DGLD builds on denoising diffusion probabilistic models [14] and the score-based/SDE formulation [15], transplants the latent-diffusion recipe [16] from images to molecules, and combines classifier-free guidance [4] with the noise-conditional classifier-guidance regime [17], injected through FiLM conditioning [18]. Unlike 3D pocket-conditioned molecular diffusion (EDM [19], GeoLDM [20]), DGLD diffuses a 1D string-derived latent, so conditioning is a function of the molecule's identity rather than of any particular conformer.

On the property side, the Kamlet–Jacobs equations [1] supply fast closed-form Tier-C detonation labels; 3D-CNN [2] and Uni-Mol [3] property predictors provide the surrogate scoring family; Politzer–Murray BDE correlations [21] and the sensitivity literature [22][23][24] ground the $h_{50}$ hazard head; thermochemical-equilibrium codes (EXPLO5 [25], Cheetah [26]) remain the absolute-value-grade reference against which our calibrated K-J estimates are scoped; and the Choi et al. review [27] codifies the field's challenges. Prior generative work on the energetic high-energy tail is scarce and label-limited [28][29]; the most direct contemporary competitor is a property-conditioned RNN coupled to QM validation [30], which differs from DGLD in prior (autoregressive SMILES vs latent diffusion), supervision (single-tier regression vs four-tier trust gating), and validation depth (QM-only vs the four-stage SMARTS → Pareto → GFN2-xTB → DFT funnel).

Evaluation and validation draw on MOSES/GuacaMol distribution-learning benchmarks [31][32] and the Fréchet ChemNet Distance [33] (read as a chemistry-class-transfer signal, since ChemNet is trained on drug-like ZINC [34]+PubChem chemistry); SA [35] and SCScore [36] as synthesisability caps and Tanimoto similarity [37] as the novelty bound; and the external validation stack GFN2-xTB [38], PySCF/gpu4pyscf [39], AiZynthFinder [40], with functional/basis choices grounded in GMTKN55 [41].

## 2. Results

### 2.1 Roadmap and headline targets

Section 2 is structured results-first: Section 2.2 reports the gated Pareto reranker top leads (Stages 1+2 of the validation chain); Section 2.3 reports physics validation (xTB triage and DFT confirmation, Stages 3+4); Section 2.4 combines novelty, retrosynthesis, and the E-set scaffold-diversity audit; Section 2.5 contrasts DGLD against no-diffusion baselines; Section 2.6 is the ablation summary. The four headline targets are $\rho \geq 1.85$ g cm$^{-3}$, $D \geq 9.0$ km s$^{-1}$, $P \geq 35$ GPa and Tanimoto novelty $\leq 0.55$ against every training row, validated through the four-stage chain SMARTS → Pareto → xTB → DFT documented in Section 4.13. The performance targets are assessed by anchor-relative ranking on a common calibrated Kamlet–Jacobs scale, on which the headline lead L1 reaches $D = 8.25$ km s$^{-1}$ and $P = 32.9$ GPa against HMX at 7.52 and PETN at 8.24 km s$^{-1}$ under the identical recipe, placing it in the HMX/PETN band. The calibrated scale is uniformly compressed relative to literature values across anchors and leads alike (literature-scale equivalents of the targets: $D \geq 9.0$ km s$^{-1}$, $P \geq 35$ GPa), so all performance claims in this paper are anchor-relative rankings on a common scale rather than absolute-value predictions (Section 2.3) (Supplementary Fig. S20). Hyperparameters and the production configuration (CFG=7, pool ≥ 40k per denoiser, alpha-anneal disabled) are documented in Section 4.15 and Supplementary Section B.5 (Supplementary Table S5) and are not re-derived here.

### 2.2 Top leads from the gated Pareto reranker (Stages 1+2)

Applying Section 4.13 gated multi-objective reranker (hard filters, banded performance score, viability, novelty, sensitivity, alerts) to the pool=40 000 candidate set: of the top-400 single-objective candidates, 45 are rejected by the hard filters (poly-nitro-on-C2, MW < 130, OB > +25 %); the 355 survivors define a Pareto front of 34 candidates (Supplementary Fig. S6). The eleven chem-pass leads are shown in Figure 2. 89/100 of the merged top-100 originate from the smallest unguided pool reranked by the Pareto scaffold composite (the rank-1 trinitro-isoxazole itself comes from a guided run); guidance acts as a high-precision lever on the top of the funnel, not as the source of bulk Pareto coverage (source-pool breakdown in Supplementary Section D.10).

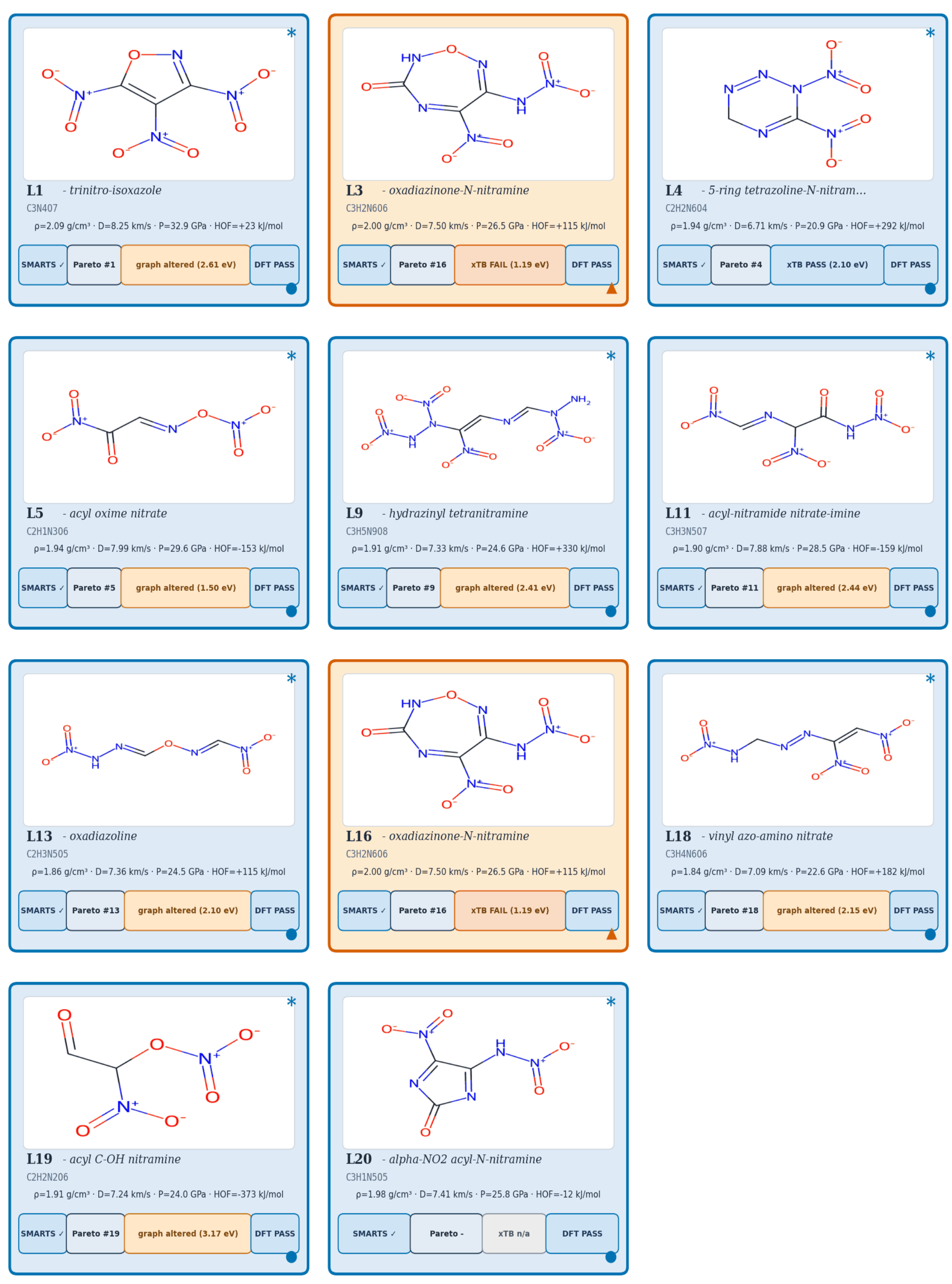

**Figure 2.** Eleven chem-pass DGLD leads (L1, L3, L4, L5, L9, L11, L13, L16, L18, L19, L20). Each card shows the RDKit 2D depiction, chemotype label, molecular formula, and 6-anchor-calibrated DFT/Kamlet–Jacobs values (ρ, D, P, HOF). The status bar reports the four validation gates: SMARTS pass, Pareto rank ("Pareto –" for L20, added from the pool=80k extension set and not assigned a top-100 rank), xTB HOMO–LUMO status, and DFT. Border and fill colour encode electronic stability: blue = xTB gap $\geq$ 1.5 eV (pass); orange = gap $<$ 1.5 eV (fail, L3/L16). The bottom-right marker is a filled circle (DFT-converged minimum) or a triangle (xTB-gap fail, L3/L16); the top-right

asterisk marks a DFT-confirmed real minimum (any "graph altered" label in the status bar is a zwitterionic-nitro canonical-SMILES round-trip artefact, not a real change).

The top leads (Figure 2, L1 and L3–L5) have viability 0.83–1.00, MW between 163 and 218 Da, and oxygen balance within ±15 %. Lead L1 is the aromatic trinitro-1,2-isoxazole, predicted at $\rho = 2.00$ g cm$^{-3}$, $D = 9.56$ km s$^{-1}$, $P = 40.5$ GPa on the Uni-Mol surrogate; its 6-anchor-calibrated Kamlet–Jacobs $D = 8.25$ km s$^{-1}$ ranks with the calibrated HMX (7.52 km s$^{-1}$) and PETN (8.24 km s$^{-1}$) anchors (Section 2.3). The Pareto front contains 34 candidates with composite $\geq 0.50$ and viability $\geq 0.83$ (Supplementary Fig. S6); the top-200 survivors populate the high-detonation corner of the $(D, P)$ plane (Supplementary Fig. S5). Unguided single-objective ranking (without the gating layer) puts a polynitro-on-C2 gaming molecule (a surrogate-gaming false positive: high predicted score, physically implausible) at the top; the gates correctly reject it (full SMARTS-trace and rejected-candidate property tabulation in Supplementary Section D.7).

### 2.3 Physics validation and DFT confirmation (Stages 3+4)

*Stage 3 (xTB triage).* The merged top-100 from Stages 1+2 is the input to Stage 3 GFN2-xTB triage at the 1.5 eV HOMO–LUMO gap gate: 85/100 survive, and 6/8 of the smaller production gated top-8 also survive. The xTB triage agrees with the Section 3 chemistry-expert critique: open-chain and strained spiro candidates fall out as low-gap, while the aromatic isoxazole and small saturated heterocycles survive cleanly. The full xTB recipe (RDKit ETKDGv3 + MMFF94 + xTB --opt tight) is in Supplementary Section C.12. Pool-size robustness: repeating the xTB triage on the gated top-15 of an unguided pool=80 000 run gives 13/15 survivors, matching the 85/100 rate of the production merged set; the 1.5 eV electronic-stability gate therefore holds its pass rate across pool size and guidance configuration. A larger unguided pool with the same gating yields the more physically-credible final set, which is the production configuration.

*Stage 4 (DFT audit).* First-principles DFT audit at B3LYP/6-31G(d) optimisation + $\omega$B97X-D3BJ/def2-TZVP single-point on the 11 chem-pass leads alongside two anchors (RDX, TATB) using GPU4PySCF (density from Bondi van-der-Waals integration with packing 0.69). All eleven chem-pass leads + two anchors are real local minima (no imaginary frequencies; min real modes 6.8–52 cm$^{-1}$, with near-zero soft-torsion modes in L9 and L20); three SMARTS-rejected reference scaffolds (R2, R3, R14) optimised under the same protocol are in Supplementary Section C.5. The candidates that do not advance to Stage 4 are accounted for in Supplementary Section C.5 (45 removed at the Stage-2 hard filters; 28 fail K-J/imaginary-frequency/composition gates).

*6-anchor calibration (Table 1).* A linear 6-anchor (RDX, TATB, HMX, PETN, FOX-7, NTO) calibration gives $\rho_{\mathrm{cal}} = 1.392\,\rho_{\mathrm{DFT}} - 0.415$ and $\mathrm{HOF}_{\mathrm{cal}} = \mathrm{HOF}_{\mathrm{DFT}} - 206.7$ kJ mol$^{-1}$, with leave-one-out RMS $\pm 0.078$ g cm$^{-3}$ on $\rho$ and $\pm 64.6$ kJ mol$^{-1}$ on HOF (full intercept derivation in Supplementary Section C.1–C.3). Calibrated densities span $\rho_{\mathrm{cal}} \in [1.84, 2.09]$ g cm$^{-3}$; L1 raw DFT $\rho_{\mathrm{DFT}} = 1.80$ calibrates to $\rho_{\mathrm{cal}} = 2.09$ g cm$^{-3}$ (the highest of the set), and its raw $\mathrm{HOF}_{\mathrm{DFT}} = +229.5$ kJ mol$^{-1}$ calibrates to $+22.9$ kJ mol$^{-1}$. This calibrated density carries the largest single uncertainty in the paper and propagates into every derived $D$ and $P$: it is a gas-phase-derived estimate, and the independent Bondi van-der-Waals packing bracket for L1 spans $\rho \in$ $[1.69, 1.87]$ g cm$^{-3}$ (Supplementary Section C.9), below the calibrated value. The applicability domain of the correction is also narrow: the six anchors span $\rho_{\mathrm{DFT}} \in [1.613, 1.689]$ g cm$^{-3}$, mapped onto an experimental range of 1.77–1.94 g cm$^{-3}$, so a short raw interval is stretched by the fitted slope of 1.392. Six of the eleven leads lie above that window and are therefore extrapolations rather than interpolations, L1 furthest at 6.6 % above the highest anchor (per-lead positions in Supplementary Table S36). Read against the literature, $\rho_{\mathrm{cal}} = 2.09$ g cm$^{-3}$ would place L1 above $\varepsilon$-CL-20 (2.04 g cm$^{-3}$), the densest CHNO explosive in practical use, which for a monocyclic tri-nitro-isoxazole is better read as an upper bound than as an expectation: the Bondi bracket and the anchor-domain excursion both point to a lower true crystal density. Crystal-structure prediction or experimental single-crystal X-ray diffraction remains the critical missing step before any density-based performance claim here can be treated as quantitative (Section 3).

*Kamlet–Jacobs recompute and headline corroboration.* Two K-J applications appear: (i) per-lead calibrated branch (Supplementary Table S30): K-J on DFT-calibrated ($\rho_{\mathrm{cal}}$, $\mathrm{HOF}_{\mathrm{cal}}$) for ranking; (ii) population residual branch (Supplementary Table S12): K-J on raw experimental ($\rho$, HOF) from 575 Tier-A labelled rows to quantify K-J under-prediction at high N-fraction. Calibrated K-J velocities span 6.71–8.25 km s$^{-1}$ across the 11 chem-pass leads (Figure 3). L1 calibrated K-J $D = 8.25$ km s$^{-1}$; the 1.31 km s$^{-1}$ residual to the Uni-Mol surrogate (9.56 km s$^{-1}$) is larger than the K-J anchor residuals in L1's own composition regime (L1: nitrogen atom fraction $f_N \approx 0.29$, OB $\approx +8$ %, placing it in PETN's K-J-reliable regime, not the high-N RDX/HMX/FOX-7 under-prediction band). Note that L1's OB sits at the upper boundary of the standard oxygen-deficient regime ($d = 2a + b/2 = 6$ vs L1's $d = 7$); we apply Kamlet–Jacobs in its unified product-distribution form (the standard treatment for mildly oxygen-rich CHNO; see Kamlet–Jacobs 1968 §III). The 1.31 km s$^{-1}$ gap is therefore plausibly Uni-Mol surrogate over-prediction in the sparsely-represented polynitroisoxazole region rather than K-J failure. The DFT-K-J recompute places L1 in the HMX-class regime by *relative ranking against anchors*; absolute $D$ values require a thermochemical-equilibrium covolume solver (Section 3). Full K-J residual decomposition (PETN/NTO chemistry; $r(f_N, \text{residual}) = +0.43$ , $p = 4 \times 10^{-27}$ on 575 Tier-A rows) is in Supplementary Section C.6–C.7. An independent Cantera ideal-gas CJ recompute over all eleven leads and all six anchors characterises product composition and energy release, and resolves a distinct axis from Kamlet–Jacobs: an ideal-gas product EOS is nearly insensitive to condensed-phase density, so the two orderings are independent (Spearman $\rho_s = -0.62$) and the recompute quantifies the covolume requirement (Supplementary Section C.13); absolute $D$ values require BKW/JCZ3 covolume corrections (Section 3), full discussion in Supplementary Section C.13.

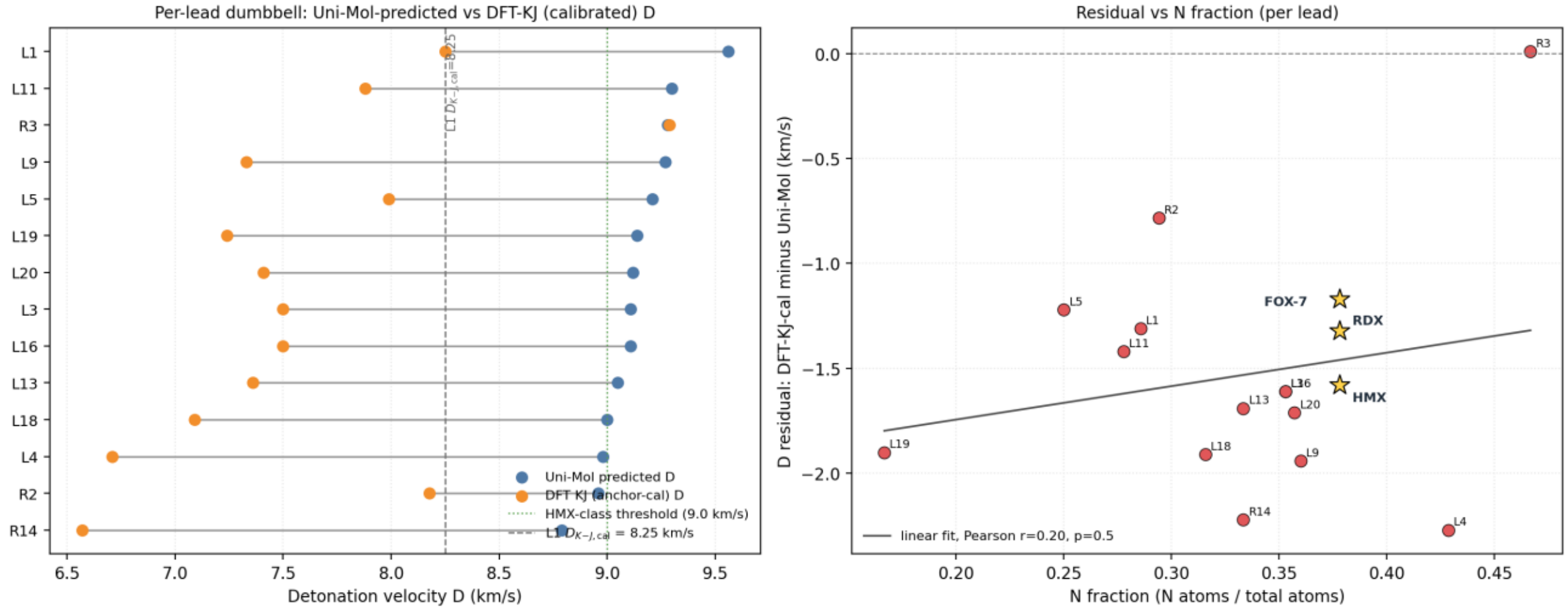

**Figure 3.** Left: dumbbell plot connecting Uni-Mol-predicted D (blue) to anchor-calibrated DFT–K-J D (orange) for each DFT-converged lead; dotted green line is the literature-scale HMX target of D = 9.0 km $s^{-1}$ (the 6-anchor-calibrated HMX anchor is 7.52 km $s^{-1}$, Section 2.3, so L1's calibrated 8.25 sits below it). Right: per-lead residual vs N-fraction with linear fit (Pearson r = 0.20 across the leads and reference compounds shown); the panel title reports the global 575-row Tier-A correlation (r = 0.43, Supplementary Table S12).

*Cross-check on SMARTS-rejected candidates.* The same DFT pipeline applied to three of the 23 SMARTS-rejected candidates (rank-2 N-nitroimine, rank-3 polyazene, rank-14 azo-amino-nitrate) finds them to be real geometric minima at B3LYP/6-31G* (min real frequencies 33–47 $cm^{-1}$). The chemist rejection is motif-level (shock-sensitivity, friction-trigger risks beyond gas-phase DFT), not geometric: the SMARTS and DFT layers are complementary, not redundant. The hazard-head post-hoc filtering recovers all 23 SMARTS rejects with perfect recall by construction and exposes 66 additional candidates on which the SMARTS catalog is silent (Supplementary Section D.12).

**Table 1.** 6-anchor DFT calibration (RDX, TATB, HMX, PETN, FOX-7, NTO). Linear corrections $\rho_{\mathrm{cal}} = 1.392\rho_{\mathrm{DFT}} - 0.415$ and $HOF_{\mathrm{cal}} = HOF_{\mathrm{DFT}} - 206.7$ kJ $mol^{-1}$; LOO RMS $\pm 0.078$ g $cm^{-3}$ and $\pm 64.6$ kJ $mol^{-1}$.

| Anchor | $\rho_{\mathrm{exp}}$ (g $cm^{-3}$) | $HOF_{\mathrm{exp}}$ (kJ $mol^{-1}$) |
|---|---|---|
| RDX | 1.82 | +70 |
| TATB | 1.94 | −141 |
| HMX | 1.91 | +75 |
| PETN | 1.77 | −538 (condensed-phase 298 K from LLNL Explosives Handbook, Dobratz 1981; literature range −504 to −539 kJ $mol^{-1}$ depending on measurement method; the chosen value is within the ±64.6 kJ $mol^{-1}$ LOO RMS) |
| FOX-7 | 1.89 | −134 |
| NTO | 1.93 | −129 |

*Literature context for L1*. The polynitroisoxazole family is established in the energetic-materials literature ([42][43]); the isomeric 3,4,5-trinitro-1*H*-pyrazole [44] is the closest fully-substituted ring previously characterised. The 3,4,5-trinitro-1,2-isoxazole isomer DGLD proposes is absent from the 65 980-row labelled master (max-Tanimoto 0.27) and from PubChem; it is therefore a chemotype-class rediscovery with a positionally novel substitution pattern.

### 2.4 Novelty, synthesisability, and scaffold diversity

Three audits frame the merged top-100 against PubChem, public USPTO retrosynthesis templates, and a scaffold-distinct E-set extension. A scaffold-diversity audit shows DGLD spans a chemotype distribution (the scaffold-distinct E-set extension, E1–E10: 10 DFT-validated candidates (7 novel, 3 known) / 8 Bemis–Murcko scaffolds / 7 families), not a single isoxazole hit.

#### *Novelty audit*

Of the merged top-100, 96/97 are absent from PubChem [45] (PUG REST on canonical SMILES; 3 transient REST errors excluded; the one rediscovery is 1-nitro-1H-tetrazol-1-amine at rank 56). Independently, 97/100 have no near-neighbour above Tanimoto 0.55 in the 65 980-row labelled master (Supplementary Table S31); of the three that do (dinitramide, 1,2-dinitrohydrazine, N,N′-dinitrocarbodiimide), one is an exact match, and all three are established high-density CHNO motifs the model recovers. Novelty here is assessed against PubChem, the SureChEMBL patent corpus, and the labelled master; all eleven leads are absent from these sources, though a full CAS Registry (SciFinder) check is recommended before a formal novelty claim. Zero of the 100 candidates lie within Tanimoto 0.70 of any augmented-corpus row, and zero are exact matches in the 694 518-row augmented corpus (Supplementary Table S31); the candidates are *more* distant from the augmented corpus than from the labelled master alone, despite the augmented corpus being >10× larger. The strengthened SMARTS catalog (N-nitroimines and open-chain polyazene/azo-nitro motifs) retains 77/100 of the merged top-100; the rank-1 trinitro-isoxazole survives (per-class breakdown in Supplementary Section D.7). Applied to the DFT-audited set, this strengthened gate flags one N-nitroimine (oxime-nitrate-imidazoline) that the production SMARTS gate passes; it is tabulated as reference R2 (Supplementary Table S8), not a chem-pass lead, so the confirmed set is 11 leads (10 unique). We distinguish two roles within this set: L1 is the single synthesis-actionable lead, being the only member for which a productive retrosynthetic route was recovered (Section 2.4), while the remaining leads are reported as a scaffold-diversity result, establishing that the sampler reaches multiple distinct energetic chemotypes rather than concentrating on one.

#### *Retrosynthesis audit*

AiZynthFinder [40] with public USPTO expansion + filter policies and the ZINC in-stock catalog (200 MCTS iter, 300 s/target) was applied to L1, L4, L5 and then extended to the remaining eight chem-pass leads. L1 returns 9 productive routes; the top route is 4 steps with state score 0.50 (Supplementary Table S32). L4, L5, and the eight extension leads return zero productive routes within budget; reproduced at 5× budget (1000 MCTS iter, 1800 s) on L4/L5. The L1 disconnection sequence (electrophilic ring-nitration; Boc protection; diphenylphosphoryl azide (DPPA)-mediated Curtius rearrangement on 4,5-dinitro-1,2-isoxazole-3-carboxylic acid), its hazard caveats (acyl-azide intermediate at primary-explosive class), and the ZINC catalog gap on energetic-domain intermediates are documented in Supplementary Section D.14. The 1/11 hit rate quantifies a public-USPTO drug-domain template-database gap, not unsynthesisability of the candidates; an energetics-domain template extension is flagged as community follow-up in Section 3.

### *E-set scaffold-diversity audit (E1–E10)*

To probe scaffold diversity beyond the single sampling stream of the L-set, we mined a 500-candidate extension pool from the four sampling runs of Section 2.2 under the same Stage 1 SMARTS gate but with a Tanimoto-NN cap of 0.55 against L1–L20. The 500 SMILES were pre-screened with GFN2-xTB on Modal CPU; 10 scaffold-distinct survivors were promoted to A100 DFT under the same protocol used for L1–L20 and post-corrected with the 6-anchor calibration. By Bemis–Murcko bookkeeping the 10 picks span 8 distinct scaffolds across 7 chemotype families: 1,2,3,5-oxatriazole (E1), NH-pyrrole nitroaromatic (E6), acyclic and small-ring nitramines (E2–E4), small-ring nitrate esters (E5, E7), geminal polynitro carbocycle (E8), bare 1H-tetrazole (E9), and nitro-imidazoline (E10); four families (oxatriazole, NH-pyrrole, acyclic nitramine, geminal polynitro carbocycle) are absent from the L-set entirely.

E1 oxatriazole: independent recovery of a known structure. E1 (4-nitro-1,2,3,5-oxatriazole) is a *known* compound (PubChem CID 57438703), not a novel structure; the unguided sampler surfaced it independently, an external check that DGLD reaches genuine energetic chemotypes. Under the same 6-anchor calibration applied to L1, E1 reaches $\rho_{cal} = 2.04$ g cm$^{-3}$, $D_{K\text{-}J,cal} = 9.00$ km s$^{-1}$, $P_{K\text{-}J,cal} = 38.6$ GPa, with a Politzer–Murray BDE-correlated $h_{50}$ of 82.7 cm; E1's calibrated $D$ exceeds L1's, though its calibrated density is marginally lower (2.04 vs 2.09 g cm$^{-3}$). The 1,2,3,5-oxatriazole ring system has known thermal/Lewis-acid ring-opening pathways (Sheremetev 2007; Katritzky 2010); a dedicated BDE and DSC/TGA stability screen is required before E1 is promoted to synthesis priority (Supplementary Section C.5.1). We therefore assign E1 a single role throughout this paper: it is an external-validation signal on the sampler's scaffold reach, not a lead. Because the 6-anchor set contains no oxatriazole-class member, its calibrated $D = 9.00$ km s$^{-1}$ is an extrapolation outside the anchor chemical space and is read as an upper bound, pending both a dedicated stability screen and an oxatriazole-class anchor extension (with a thermochemical-equilibrium CJ recompute on calibrated inputs, scoped as future work in Section 3). It is not counted among the synthesis-actionable results.

Of the 10 E-set candidates (Supplementary Tables S33 and S34), 5 of 9 with defined K-J clear $D_{K\text{-}J,cal} \geq 7.0$ km s$^{-1}$ (E1–E5); 8 of 10 have $h_{50BDE} \geq 30$ cm. E9 (bare 1H-tetrazole) is a deliberate filter-check: K-J is undefined at its oxygen balance and Stage 4 correctly leaves $D$ and $P$ unreported. E2's K-J $D$ (9.22 km s$^{-1}$) is upper-bound only because its OB = +28.9 % exceeds the +25 % K-J reliability limit (Supplementary Section D.13). The L-set sits at the upper edge of a credible distribution; L1 is the top-ranked novel lead, and the sampler's independent recovery of the known oxatriazole E1 shows its reach spans chemically distinct families.

## 2.5 Comparison with no-diffusion baselines

Four no-diffusion baselines were run through the same downstream pipeline on the same training corpus to isolate the contribution of the diffusion prior and guidance. The Gaussian-latent control and the matched-compute guided-vs-unguided headline are reported in Supplementary Section F.2 and Supplementary Section F.3 respectively. Results are in Table 2 and Supplementary Fig. S7.

**Table 2.** Top-1 comparison: SMILES-LSTM, MolMIM 70 M, SELFIES-GA, and REINVENT 4 no-diffusion baselines vs. unguided and best-novel DGLD conditions (head-to-head in Supplementary Section F.4). Columns: top-1 composite; $D$, $\rho$, $P$; max-Tanimoto to LM; seeds; memorisation rate (fraction of valid unique samples that are exact labelled-master matches). †SELFIES-GA composite is internal-to-the-run (viability fixed at 0.5, novelty assumed 1.0 during search); it is not on the same basis as the DGLD/LSTM pipeline composite and is omitted from direct comparison. ‡REINVENT 4 composite is N-fraction proxy (not full DGLD reranker); $D/\rho/P$ are Uni-Mol scores for the seed-42 top-100 by N-fraction, post-hoc (the RL reward was N-fraction, not the DGLD composite); max-Tanimoto for the top-100 is in the [0.15, 0.65] novelty window (not memorisation).

| **Method/condition** | **top-1 composite *(lower = better)*** | **top-1 $D$ (km s$^{-1}$)** | **top-1 $\rho$ (g cm$^{-3}$)** | **top-1 $P$ (GPa)** | **top-1 max-Tani to LM** | **seeds** | **memo rate** |
|---|---|---|---|---|---|---|---|
| SMILES-LSTM (no diffusion) | 0.083 | 9.58 | 1.96 | 40.0 | **1.000** (exact LM match) | 3 | **18.3% ± 0.5%** |
| MolMIM 70 M (drug-domain pretrain, no diffusion) | 4.79 | 7.70 | 1.76 | 25.5 | 0.625 | 1 | n.d. |
| SELFIES-GA (property optimisation, 2 000 pool, 30 gen)† | n.c.† | 9.54 | 1.994 | 40.9 | **1.000** (exact LM match; 75/100 rediscoveries) | 1 | 75% |
| REINVENT 4 (N-frac RL, 40k pool, seed 42)‡ | 0.42‡ | 9.02‡ | 1.85‡ | 34.5‡ | 0.57 (amino-tetrazine); 0.32–0.38 (seeds 1-2) | 3 | **near-zero** <0.1% (0.04% exact match, seeds 1-2; <1% novelty-window, seed 42) |
| DGLD Hz-C0 = SA-C0 unguided (cfg-only) | 0.451 ± 0.126 | 9.44 ± 0.07 | 1.93 ± 0.01 | 39.7 ± 0.6 | 0.61 ± 0.10 | 6 | 0% |
| DGLD Hz-C2 viab+sens+hazard | 0.485 ± 0.152 | 9.39 ± 0.04 | 1.91 ± 0.03 | 38.7 ± 0.6 | **0.27 ± 0.03** | 3 | 0% |

*SELFIES-GA collapses under DFT audit.* SELFIES-GA (2k-molecule pool, 30 generations) returns 75/100 top candidates as exact corpus rediscoveries (top-1 is a rediscovery); the best novel candidate at 2k is rank 5 ($D$ = 9.39 km s$^{-1}$, max-Tanimoto 0.487); at 40k pool, the best novel outlier reaches $D_{surrogate}$ = 9.73 km s$^{-1}$ but collapses to $D_{DFT}$ = 6.28 km s$^{-1}$ under the same DFT audit chain applied to DGLD leads (3.5 km s$^{-1}$ surrogate artefact; the Uni-Mol is not calibrated for nitro-oxadiazole/triazole fused scaffolds). SMILES-LSTM (2-layer, 6 M parameters, trained on the same 326 k SMILES corpus): top-1 is an exact labelled-master rediscovery (max-Tanimoto =

1.000); memorisation rate 18.3% ± 0.5% across 3 seeds, seed-stable. The best novel top-1 is a 5-atom aminotriazole fragment with no Uni-Mol score; the model reproduces training data, not new energetic leads. MolMIM 70 M (drug-domain pretrained): top-1 novel at Tanimoto 0.625 but at $D = 7.70$ km s$^{-1}$, far below HMX; uncalibrated for the energetic regime. REINVENT 4 (N-fraction RL reward, 3 seeds, 40k pool): generates genuinely novel high-N heterocycles with exact memorisation below 0.1%; seed-42 top-100 Uni-Mol-scored at top-1 $D = 9.02$ km s$^{-1}$, 0.37 km s$^{-1}$ below DGLD Hz-C2. N-fraction RL is a useful novelty lever but does not optimise directly for the D/ρ/P targets DGLD conditions on. DGLD Hz-C2 is the only condition with consistent novel productive-quadrant coverage confirmed at DFT level. Supplementary Section E lists the ten most-novel candidates from each baseline pool.

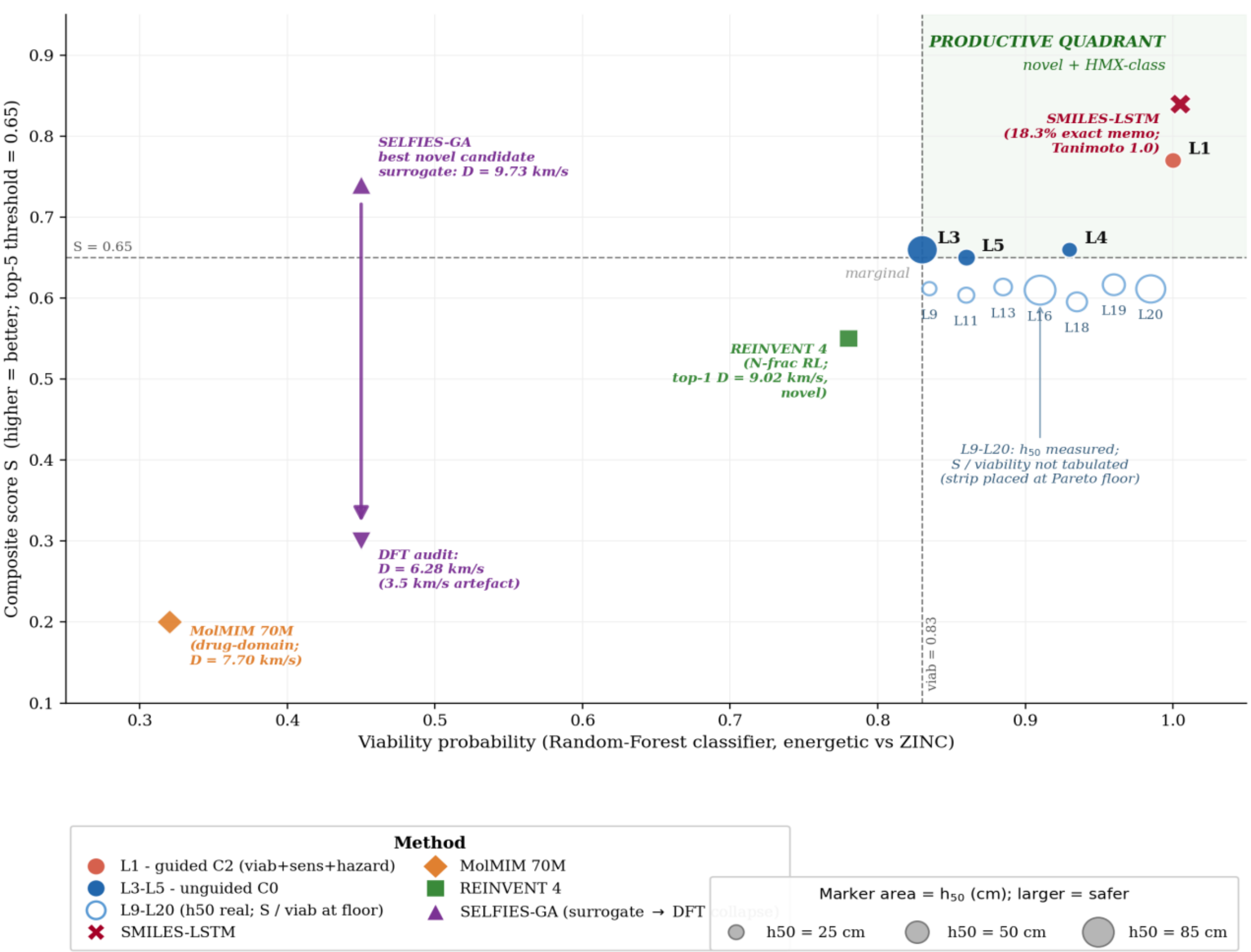

**Figure 4.** Productive-quadrant scatter for the 11 DFT-confirmed lead cards (10 unique) with the four no-diffusion baselines as reference markers. $x$-axis: viability probability (RF classifier, energetic vs ZINC); $y$-axis: composite score $S$ (higher = better). Dashed lines mark the top-5 thresholds ($S = 0.65$, viab $= 0.83$); the green-tinted upper-right quadrant is the productive zone (novel + HMX-class). Marker area is proportional to drop-weight impact sensitivity $h_{50}$ (cm; larger =

safer). L1 (guided Hz-C2, red) sits at the top of the productive quadrant; L3–L5 (filled blue, unguided C0) cluster near the threshold boundary; L9–L20 (hollow blue) carry real $h_{50}$ but $S$ and viability are not tabulated and are placed along the Pareto-floor strip. Baseline markers ($S$ not on the DGLD composite scale; placed as reference): SMILES-LSTM (red X, memorised; Tanimoto 1.0), MolMIM 70 M (orange diamond, drug-domain; $D = 7.70$ km s$^{-1}$), REINVENT 4 (green square, N-fraction RL; top-1 $D = 9.02$ km s$^{-1}$, novel), SELFIES-GA (purple, surrogate → DFT collapse: top-1 falls from $D = 9.73$ to 6.28 km s$^{-1}$ under DFT audit). Source: $S$/viab for L1, L3–L5 from Section 2.2 prose; $h_{50}$ from Supplementary Table S7; baselines from Table 2.

**Note on Figure 4 score conventions.** The composite score $S$ on the $y$-axis is the higher-is-better Stage-1+2 reranker success score (range 0–1), not the lower-is-better Pareto-reranker penalty tabulated in Table 2, Supplementary Fig. S7, and Supplementary Table S27. Cross-figure ranking is consistent (DGLD Hz-C2 best on both) but absolute values are not comparable.

*Distribution-learning metrics.* SMILES-LSTM has FCD = 0.52 against a 5 000-row labelled master sample (distributionally indistinguishable because it reproduces labelled rows); DGLD has FCD = 24–26 across guidance conditions: the diffusion sampler performs a *targeted search off the prior*, not corpus mimicry, with anti-correlation between FCD and Pareto-reranker composite within DGLD confirming the design intent. Full small-multiples (validity, scaffold uniqueness 659–1262, IntDiv1 0.818–0.838) are shown in Supplementary Fig. S8, with per-condition values in Supplementary Section D.11.

### 2.6 Ablation summary

Seven ablations measure the contribution of each system component to the headline. Supplementary Table S35 lists the headline result of each; full prose, sub-tables, and figures are in Supplementary Section F. The forest plot of effect sizes (Supplementary Fig. S9) summarises the guidance-axis ablations visually.

*What each component contributes.* The tier-gate is the single largest contributor: removing it collapses the sampler to high-N degenerate open chains (53.9 % keep-rate, no ring chemistry) rather than the ring-bearing high-density manifold the production model occupies. The diffusion prior contributes a +0.45 km s$^{-1}$ $D$-lift over a compute-matched Gaussian-latent control, whose 48 % vs 4.6 % Stage-1+2 keep-rate reversal confirms that the diffusion prior's role is not chemical-plausibility filtering (a Gaussian decode through frozen LIMO does that easily) but distribution concentration on the high-$D$ tail. Multi-head guidance trades scaffold diversity for novelty: hazard-axis C2 reaches max-Tanimoto 0.27 (the most novel condition) at the cost of scaffold count 5 vs 12. Pool fusion across 5 lanes lifts post-filter yield nearly five-fold (4 639 vs 966 candidates, +5×) without lifting top-1, surfacing 24 Bemis–Murcko scaffolds vs 7.

*Seed-variance context for the headline.* Per-condition seed variance (3-seed s.d. on top-1 composite, range 0.106–0.184) is comparable to the across-condition mean differences, so the production C2 default is justified by the most-novel result, with top-1 $D$ invariant across conditions; the qualitative ablation conclusions are robust to seed across all six guidance axes (Supplementary

Table S27, full multi-seed table). The 6-seed unguided baseline gives the tightest mean (top-1 composite 0.451 ± 0.126); the SA-axis SA-C3 (viab+sens+SA) is the worst at 0.698 ± 0.015.

*What is ruled out.* A drug-domain SA-gradient head (RDKit SAScorer, calibrated on Reaxys/pharma reaction corpora) worsens the composite by 13 % in our energetic-materials regime; production therefore uses $s_{\text{SA}} = 0$. The SA and SC heads serve as multi-task trunk regularisers rather than sample-time gradients; the steering bus invokes the three domain-native heads. Both heads add ~2 × 256k parameters of trunk-regularisation supervision during training (Supplementary Section B.4 contains the full SA / SC drug-domain transfer-head story).

## 3. Discussion

DGLD's central result is that tier-gated training and switchable sample-time steering together make a latent diffusion model a *productive-quadrant generator*: the tier gate keeps miscalibrated labels out of the conditional gradient, so sampling concentrates where candidates are both novel and, under first-principles validation, competitive with the HMX/PETN reference class. The top-ranked novel lead L1 (3,4,5-trinitro-1,2-isoxazole) and the independently recovered known oxatriazole E1 (4-nitro-1,2,3,5-oxatriazole, PubChem CID 57438703) come from disjoint chemotype families on a single sampling run, external evidence that DGLD's scaffold reach spans multiple chemotype classes; the 10 novel L-set leads themselves cover several scaffolds (E1's placement pending an oxatriazole-class DFT anchor). The mechanism that makes this possible is the label-trust gate: routing only the ~3,000 rows that carry a trustworthy experimental or DFT label on the target detonation channels into the conditional gradient, while the remaining lower-confidence labels train only the unconditional prior, prevents miscalibrated Kamlet–Jacobs and surrogate labels from steering generation toward physically implausible high-scoring regions, the failure mode visible in the SELFIES-GA baseline, whose best novel candidate loses 3.5 km $s^{-1}$ under DFT audit.

Relative to prior work, DGLD occupies a distinct position. Discriminative property surrogates score candidates but do not propose them; generative language models trained on energetic corpora tend to memorise, as the SMILES-LSTM baseline's 18.3% exact-rediscovery rate shows; and standard classifier guidance degrades over the short trajectories molecular generation requires. By coupling generative inverse design to a graded validation funnel that ends in density-functional theory, DGLD converts a sparse-label liability into a usable training signal and returns a ranked, physics-checked candidate list rather than an unvalidated pool. Because the gating mechanism depends only on the availability of a trust hierarchy over labels (not on any energetic-materials-specific structure), the recipe should transfer to other data-limited inverse-design settings, with only the validation funnel changing per domain.

*Limitations. Crystal packing is the dominant unquantified error source.* All densities are estimated from gas-phase DFT geometry using Bondi van-der-Waals volumes with a fixed packing factor of 0.69. This factor varies from approximately 0.65 (loosely packing aromatics) to 0.72 (cubane-class compounds), with literature reports up to 0.78 for the densest CHNO crystals (Supplementary Section C.3); a ±5% packing-factor error alone propagates to ±0.10 g $cm^{-3}$ in density, which at the K-J sensitivity of $\partial D / \partial \rho \approx 2.9$ km s$^{-1}$ per g cm$^{-3}$ (Supplementary Table S30) yields ±0.29 km s$^{-1}$ in $D$, comparable to the 6-anchor calibration uncertainty of ±0.23 km s$^{-1}$. Note that this propagation applies to the *uncalibrated* Bondi-vdW estimate; the 6-anchor empirical calibration absorbs most of the systematic packing-factor error and brings the residual to ±0.078 g cm$^{-3}$ (Supplementary Section C.4). Polymorph screening is absent; high-nitrogen heterocycles frequently exhibit multiple crystal forms with different densities and sensitivity profiles (cf. $\varepsilon$- vs other CL-20 polymorphs). Crystal structure prediction or experimental single-crystal X-ray diffraction is the critical missing step before any density-based performance claim can be considered quantitative. Crystal structure prediction is the natural follow-up validation; recent benchmarks of CSP on energetic molecular crystals [46] demonstrate that polymorph screening is feasible at this candidate-list scale.

*Independent cross-checks on L1 and E1.* Two independent semi-empirical cross-checks confirm conservative bounds. A Bondi-vdW packing-factor bracket on L1 and E1 (Supplementary Section C.9) yields $\rho \in [1.69, 1.87]$ g cm$^{-3}$ for L1 and [1.65,1.83] g cm$^{-3}$ for E1, both below the 6-anchor-calibrated values (2.09 and 2.04); the 14% offset reproduces the calibration slope and confirms the pure Bondi-vdW estimator is conservative. A coordinate-preserving GFN2-xTB BDE scan (Supplementary Section C.10) places the weakest-bond cleavage of L1 at 86 kcal mol$^{-1}$ and E1 at 93 kcal mol$^{-1}$, both on C–$NO_2$ nitro-loss channels with no sub-80 kcal mol$^{-1}$ channel that would predict primary-explosive sensitivity.

The Uni-Mol surrogate is well-calibrated on the labelled distribution but extrapolates onto the high-density tail with unquantified reliability; the top leads should be treated as candidates for DFT and experimental validation, not as final answers. SA and SCScore are heuristic synthesisability bounds. Absolute $D$ values are reported under the closed-form Kamlet–Jacobs approximation; absolute-value-grade numbers require a thermochemical-equilibrium CJ code with a covolume EOS (EXPLO5, Cheetah-2, or the open-source Cantera-based Shock and Detonation Toolbox), which we do not run in this work. The six DFT anchor compounds contain no oxatriazole-class member; E1's $D_{\text{K-J,cal}} = 9.00$ km s$^{-1}$ is therefore an extrapolation outside the anchor chemical space. DNTF lies outside the calibration's applicability domain (a furazan-class outlier, Supplementary Section C.11), so the 6-anchor calibration is retained; an oxatriazole-class anchor with experimental crystal density and detonation data remains scoped as future work before E1 reaches L1 confidence. L1's $\rho_{\text{cal}} = 2.09$ g cm$^{-3}$ also involves nitroisoxazole chemotype extrapolation: a packing factor of 0.65 (lower-end aromatic, vs 0.69 used) would give $\rho_{\text{cal}} \approx 1.94$ g cm$^{-3}$ on the calibrated scale, equivalently the raw Bondi value $\rho \approx 1.69$ g cm$^{-3}$ quoted in Supplementary Section C.9 before the 6-anchor calibration is applied, shifting $D_{\text{K-J}}$ by roughly $\pm 0.3$ km s$^{-1}$.

The 1.5–2 km s$^{-1}$ Kamlet–Jacobs vs Uni-Mol residual reported in Section 2.3 is consistent with the typical disagreement between Kamlet–Jacobs and full Chapman–Jouguet thermochemical-equilibrium codes (EXPLO5, Sućeska et al.; Cheetah-2, Fried–Howard) on high-N CHNO compounds: published benchmarks place this disagreement in the 0.3–1.5 km s$^{-1}$ band for typical organic explosives and at the higher end (1–2 km s$^{-1}$) for compounds with N-fraction $\gtrsim 0.4$, where the K-J fixed-product-distribution assumption breaks down. The population-level Section 2.3 evidence (Pearson $r(f_N,\text{residual}) = +0.43$, $p < 10^{-26}$) is itself a regime-aware reproduction of the same effect on 575 Tier-A experimental rows. Definitive absolute-$D$ numbers for the top lead therefore require either a thermochemical-equilibrium CJ recompute (EXPLO5, Cheetah-2, or the open-source Cantera SDT) on the calibrated DFT inputs, or experimental synthesis; the Uni-Mol $D$ should be read as relative-ranking-grade, not absolute-value-grade.

*Public USPTO retrosynthesis templates are drug-domain-biased for energetic-materials chemistry.* AiZynthFinder finds a 4-step productive route for L1 (state score 0.50) but not for the ten remaining lead cards, which reflects the Reaxys-/USPTO-derived template database rather than unsynthesisability of the candidates; energetic-domain templates (nitration, $N_2O_5$ nitrolysis, ring-cyclisation) would be required for an informative retro-screen.

The statistical scope of the headline numbers is bounded as follows. Supplementary Section F.4 reports a 4-condition × 3-seed head-to-head at pool=10k with a no-diffusion SMILES-LSTM baseline, and Supplementary Section F.3 reports the single-seed pool=40k head-to-head used for the merged top-100; tables elsewhere use a single sampling seed. 89/100 of the merged top-100 come from a single unguided pool reranked by the Pareto-reranker scaffold composite, so the headline novelty/stability numbers reflect this single sampling stream, with the multi-seed Supplementary Section F.4 matrix as the variance estimate. Each denoiser is a ∼6 hr training run; the v4b production architecture is trained at three seeds and the production C2 condition (Hz-C2 viab+sens+hazard, pool = 10k per seed) resampled. Across the 3 denoiser seeds, top-1 $D$ = 9.38 ± 0.01 km s$^{-1}$ (relative s.d. 0.1%), top-1 $\rho$ = 1.92 ± 0.01 g cm$^{-3}$, top-1 $P$ = 38.8 ± 0.07 GPa. The denoiser-init seed variance is far smaller than the across-condition C0 → C2 spread (∼1% of mean), so the production C2 lift is stable across denoiser-init seeds.

The diffusion model trains on the full 694k augmented corpus; the candidate-discovery objective is to extrapolate off the labelled distribution rather than recover held-out labels, so we do not report a held-out test split. The Tanimoto-stratified novelty against the 694k corpus (Section 2.4: median 0.32, 0/100 within Tanimoto 0.70) is the train-leakage proxy. The Uni-Mol surrogate → xTB triage → DFT minimum chain is not a substitute for first-principles equilibrium thermochemistry or experimental synthesis; we position the paper as a methodology and a candidate-list paper, not a synthesis paper. Finally, the cfg-dropout (0.10), property-dropout (0.30), high-tail oversample (10×), and the Tanimoto window (0.20, 0.55) are fixed configuration; no full hyperparameter grid is reported (Supplementary Section D.3).

*Conclusions.* DGLD demonstrates that a label-quality gate at training time, multi-head score-model guidance at sample time, and a four-stage chemistry-validation funnel are sufficient to produce novel CHNO leads in the HMX/PETN performance band on commodity hardware. Eleven lead cards (10 unique compounds; L3 and L16 coincide) are DFT-confirmed local minima; the headline compound (L1, 3,4,5-trinitro-1,2-isoxazole) reaches $\rho_{\text{cal}} = 2.09$ g cm$^{-3}$ and $D_{\text{K-J,cal}} =$ 8.25 km s$^{-1}$ and is structurally dissimilar from all 65 980 training molecules (nearest-neighbour Tanimoto 0.27). Three strong energetics-domain baselines fail the novel-and-on-target test on the same compute: SMILES-LSTM memorises 18.3% of outputs; SELFIES-GA sees its best novel candidate collapse to $D_{\text{DFT}} = 6.28$ km s$^{-1}$ under DFT audit; REINVENT 4 generates novel high-N heterocycles but peaks at $D = 9.02$ km s$^{-1}$. Across all four baselines, among the baselines only SELFIES-GA's top candidate was DFT-audited (and collapsed); DGLD is the only method with productive-quadrant coverage confirmed at DFT level. (Kamlet–Jacobs relative ranking throughout; absolute $D$ requires a thermochemical-equilibrium solver, Section 3.)

On the same sampling run the sampler recovered the known oxatriazole E1 (Section 2.4), whose calibrated $D_{\text{K-J,cal}} = 9.00$ km s$^{-1}$ comes from a scaffold family disjoint from L1; its recovery validates the sampler's scaffold reach rather than adding a novel lead, and its calibrated D is an upper bound pending thermal stability confirmation and a 1,2,3,5-oxatriazole-class DFT anchor (Section 3).

*Recommendation for synthesis-and-characterisation.* L1 meets five criteria for an experimental campaign: it is absent from PubChem, SureChEMBL patent structures, and the 65 980-row labelled master, AiZynthFinder finds a 4-step productive route (the energetic-domain intermediates are flagged in_stock: false against ZINC, Section 2.4; subject to the Supplementary Section D.14 acyl-azide intermediate hazard noted there: DPPA Curtius rearrangement, the acyl-azide is a primary-explosive-class motif requiring careful handling), the GFN2-xTB HOMO–LUMO gap is 2.6 eV, the GFN2-xTB weakest-bond BDE is 86 kcal mol$^{-1}$ on a C–$NO_2$ bond (the dominant decomposition channel; nitro loss, with no sub-80 kcal mol$^{-1}$ primary-explosive sensitivity flag), and DFT confirms a real local minimum. Polymorph screening and crystal-density refinement (Section 3) are required before any density-derived performance claim can be considered quantitative.

Three extensions would close the remaining gaps. First, replace the closed-form Kamlet–Jacobs step with a thermochemical-equilibrium covolume CJ solver. The open-source Cantera-based Shock and Detonation Toolbox is the practical path; EXPLO5 and Cheetah-2 are commercial alternatives. Either choice, applied to the 6-anchor-calibrated DFT inputs, converts relative-ranking-grade predictions into absolute-value-grade ones and removes the K-J caveat from the headline. Second, an energetics-domain retrosynthetic template library covering nitration, $N_2O_5$ nitrolysis, and ring-cyclisation reactions would close the 1/11 USPTO-template hit rate observed here, making the synthesis-plausibility check informative across the full lead set. Third, an active-learning loop closing the diffusion-sampler / first-principles-audit cycle would continuously expand the high-tier label pool from inside the system rather than from external curation. The full pipeline (LIMO checkpoint, two conditional latent denoisers, two multi-head score models, the 4-stage validation funnel, and the 918 mined hard negatives) is released as a runnable code bundle on Zenodo (see the Data Availability statement), so the next compound to enter the HMX-class band can be discovered, validated, and recommended for synthesis at the cost of a few GPU-days.

## 4. Methods

*Dataset construction (Section 4.1–4.3) and the DGLD model, sampler, and validation funnel (Section 4.4 onward) are described below; full per-component hyperparameters and the DFT protocol are in the Supplementary Information.*

We assemble the training data by collecting molecules from several public sources that differ in both reliability and chemical scope. A small experimental core carries detonation properties measured in the laboratory; a larger DFT-quality slice carries energies and densities from first-principles simulation; a wider Kamlet–Jacobs-derived layer carries closed-form estimates from quoted density and heat of formation; and a model-derived majority carries surrogate predictions from a Uni-Mol ensemble trained on the union of the higher tiers. Some sources are restricted to energetic CHNO and CHNOClF chemistry (a labelled master of ~ 66 k canonical SMILES (note: the labelled master includes CHNOClF molecules; the Stage-1 SMARTS gate at output removes all halogens, so all reported leads are CHNO-only) with at least one measured property and a motif-augmented expansion that systematically substitutes nitro / nitramine / azide / tetrazole groups onto labelled scaffolds); others draw from broader chemistry corpora (an unlabelled energetic SMILES dump of ~ 380 k from the same chemical domain) that the diffusion model uses only to learn the unconditional prior. Every row is canonicalised to a single RDKit canonical SMILES, deduplicated across sources, and tagged with a trust tier; each property label is then used according to its reliability via the conditioning-mask recipe of Section 4.1. After cross-source deduplication and token-length filtering, the merged pretraining pool comprises ~ 694 k unique molecules (pre-deduplication: 66k labelled + 380k unlabelled CHNO + 1.08M motif-augmented expansions = 1.53M; after canonical-SMILES deduplication across all three pools: 694k unique molecules; the motif-augmented pool contributes ~230k survivors after dedup with the base two pools), of which ~ 66 k carry at least one property label. The complete source provenance (per-source row counts, citations, license terms, role in the four-tier hierarchy), the canonicalisation and tokenisation pipeline, and the high-tail oversampling ratios used during training are documented in Supplementary Section A.1 / D. Joint property distributions and per-property histograms over the labelled corpus are shown in Supplementary Figs S10 and S11; the label-tier composition (Supplementary Fig. S12) sits alongside the four-tier table in Section 4.1; atom-composition statistics over a 30k subsample are shown in Fig S1.

### 4.1 Four-tier label hierarchy

Available property labels in the energetic-materials literature span four orders of reliability, from a small core of experimental measurements to a large majority of model-derived estimates. Concatenating all tiers would let the noisiest tier dominate the conditional gradient and degrade calibration in the high-property tail; discarding the noisier tiers would throw away most of the training signal. We resolve this by annotating each property label with a trust tier and gating the conditional gradient accordingly: only Tier-A and Tier-B rows participate in the conditional signal, while Tier-C and Tier-D rows train the unconditional prior via classifier-free-guidance dropout (Section 4.6). Each label per row carries one of the following tiers (Table 3):

**Table 3.** Four-tier label-trust hierarchy. Tier-A (experimental) and Tier-B (DFT/thermochemical-derived) drive the conditional gradient via FiLM modulation; Tier-C (Kamlet–Jacobs derived) and Tier-D (ML surrogate) train the unconditional prior via classifier-free-guidance dropout (Section 4.6). Counts give the number of master rows carrying at least one label at that tier; a row can carry labels at several tiers across different properties, so the counts do not partition the 65 980-row master. Tier-A is dominated by experimental crystallographic densities [47]; only ~3 000 rows carry a trusted (Tier-A/B) label on the target detonation channels ($D$, $P$, HOF) that steer conditioning.

| Tier | Source | ~ rows | Trusted for conditioning? |
|---|---|---|---|
| A | Experimental measurement (crystallographic densities; measured detonation properties) | ~19 200 | yes |
| B | DFT/thermochemical-derived (HOF, $D$, $P$) | ~1 900 | yes |
| C | Kamlet–Jacobs derived from quoted density & HOF (regime-limited: reliable for oxygen-deficient CHNO with $2a + b/2 \geq d \geq b/2$; under-predicts $D$ for high-N low-H compounds where the assumed gas-product distribution is wrong) | ~12 000 | no (mask out) |
| D | ML surrogate (Uni-Mol smoke ensemble; generative-model predictions) | ~47 000 | no (mask out) |

The denoiser conditional gradient fires only on Tier-A and Tier-B rows; Tier-C and Tier-D rows train the unconditional prior via classifier-free-guidance dropout (Section 4.6). This recipe lets the full corpus shape the marginal density of the generative model without letting noisy labels contaminate the conditional gradient. Each row carries a graded *tier weight* $\omega \in [0,1]$, with default values $w_A = 1.0$, $w_B = 0.7$, $w_C = 0.3$, $w_D = 0.1$; later sections gate the conditional gradient and weight the per-row loss by these values.

### 4.2 Canonicalisation and tokenisation

Every SMILES string is canonicalised with RDKit [48] (a fixed traversal order so that the same molecule yields the same string regardless of how the source database wrote it), converted to SELFIES (a robust tokenisation in which every legal token sequence decodes to a chemically valid molecule), and tokenised to the LIMO alphabet. Tokenisation, alphabet size, and the per-source preprocessing pipeline are in Supplementary Section A.2. This step is purely data-side: no learned parameters are fit here; the output is a token-tensor representation of the corpus that the model of Section 4 will consume.

### 4.3 Training-time high-tail oversampling

The labelled detonation-velocity distribution is sharply peaked around 7–8 km s$^{-1}$, so during training Tier-A/B rows above the 90th percentile are oversampled by 5×–10×; this *asymmetric high-tail oversampling* raises the mean predicted velocity of the top leads by 0.4 km s$^{-1}$ relative to a uniform-sampling control. The oversampling ablation is in Supplementary Section D.4. The two denoiser variants of Section 4.6 differ in which property tail they apply this oversampling to (HOF tail vs ρ/D/P tail).

## 4.4 Overview and pipeline map

DGLD is a four-stage pipeline (Figure 5; per-stage panels in Supplementary Figs S13–S21): an encode-once LIMO VAE (Supplementary Fig. S13) feeds a conditional latent DDPM denoiser (Supplementary Fig. S15), whose sampling trajectory (Supplementary Fig. S19) is steered by a multi-task score model trained offline (Supplementary Fig. S17), and whose decoded candidates are filtered by a SMARTS+Pareto rerank funnel (Supplementary Fig. S20). The rest of Section 4 is organised one subsection per diagram panel; small panels are paired (Section 4.9 covers Supplementary Fig. S16 and Section 4.10 covers Supplementary Fig. S17), and the post-decode pipeline is split across three subsections (Section 4.12 Sampling covers Supplementary Fig. S19, Section 4.13 Filtering covers Supplementary Fig. S20, Section 4.14 Pool fusion covers Supplementary Fig. S21). Section 4.15 documents how every numeric constant in Section 4.5–Section 4.14 was selected. LIMO encoding is run once and cached, so all training and sampling happens in latent space; the decoder is invoked only at the end to read candidates back as SMILES. The conditional gradient is restricted to Tier-A and Tier-B rows; Tier-C and Tier-D rows drive the unconditional CFG branch only (Section 4.1, Section 4.6).

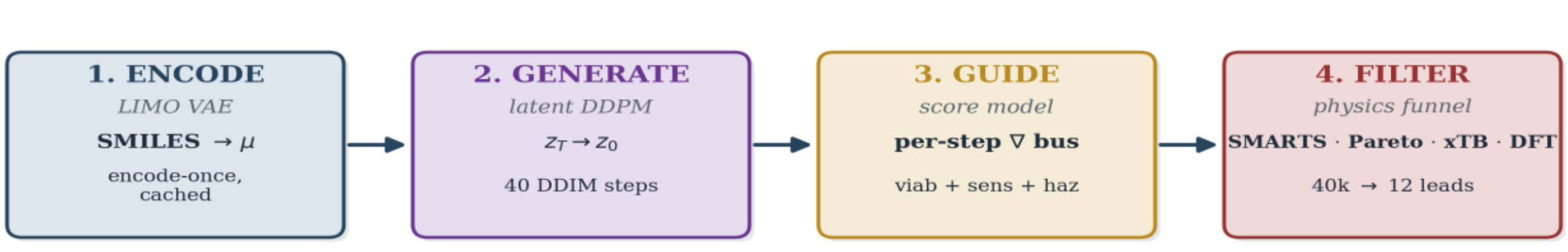

**Figure 5.** DGLD pipeline preview: encode (LIMO VAE) -> generate (conditional latent DDPM) -> guide (multi-task score model) -> filter (SMARTS, Pareto, xTB, DFT). The trust-gating annotation under the row reminds the reader that Tier-A/B labels drive the conditional gradient while Tier-C/D drive the unconditional CFG branch only. Stage references inside each box point at the per-stage panel that walks it.

## 4.5 LIMO fine-tuning and one-time latent caching

LIMO is the molecular-text VAE that supplies the cached latent $\mu$ used everywhere downstream. We fine-tune the pretrained checkpoint on the 326k energetic corpus to specialise the encoder, then freeze it for the rest of the pipeline. Sampling during fine-tuning is uniform; high-tail oversampling lives at Section 4.7 denoiser-training stage and is not applied here.

We start from the pretrained LIMO checkpoint distributed with [9]. The encoder maps a (B, 72) SELFIES-token tensor through a 64-dim embedding and a four-layer MLP (Linear(72·64→2000)–ReLU–Linear(2000→1000)–BN–ReLU–Linear(1000→1000)–BN–ReLU–Linear(1000→2·1024)) to a (B, 1024) Gaussian latent (the final 2·1024 head packs $\mu$ and $\log \sigma^2$); the decoder mirrors this with Linear(1024→1000)–BN–ReLU–Linear(1000→1000)–BN–ReLU–Linear(1000→2000)–ReLU–Linear(2000→72·108) and produces a (B, 72, 108) log-probability tensor in parallel (no autoregression). Layer-by-layer widths and parameter counts are in

Supplementary Table S2. We fine-tune all parameters for ∼8.5k steps on the 326 k energetic-biased SMILES corpus with the standard evidence lower bound (ELBO),

$$\mathcal{L}(x) = -\mathbb{E}_{q(z|x)}[\log p(x|z)] + \beta \cdot \mathrm{KL}(q(z|x) \parallel \mathcal{N}(0, I)),$$

with $\beta = 0.01$ and free-bits clipping disabled. After fine-tuning, validation token-accuracy is 64.5 %. The LIMO decoder is non-autoregressive (parallel decode); SELFIES syntax guarantees token-level validity, so molecule-level (full-sequence) validity is 100% by construction. Reconstruction accuracy (fraction of latent round-trips that reproduce the input SMILES exactly) is 31.4% on the energetic validation set, consistent with the expected LIMO performance in this domain. The latent posterior is concentrated: $\parallel \mu \parallel \approx 8$ on average, well below the $\sqrt{1024} \approx 32$ expected of $\mathcal{N}(0, I)$.

After fine-tuning the encoder is frozen and every row of the data corpus (Section 4) is passed through it once to produce a deterministic latent mean $\mu \in \mathbb{R}^{1024}$. The cached tensor stores, per row, the latent $\mu$, a property matrix (the four conditioning targets $\rho$, HOF, $D$, $P$), a tier matrix, a per-row trust mask, and per-property normalisation statistics. The encoder posterior variance is discarded by design; $\mu$ is treated as a deterministic anchor in latent space, and no SMILES is ever re-encoded at sample time. The denoiser of Section 4.6 and the score model of Section 4.10 read from this cached tensor only.

### 4.6 Conditioning mask construction

Each gradient step samples a fresh per-row mask $m \in \{0,1\}^4$ that decides which of the four conditional properties are exposed to the denoiser at that step. The mask is the mechanism by which tier-eligibility (which labels a row carries) interacts with classifier-free dropout (the unconditional branch trained alongside the conditional).

The per-step training mask $m \in \{0,1\}^4$ controls which conditioning properties enter FiLM at each step; the construction is a 5-stage stochastic pipeline over the per-row eligibility $e$ and tier weight $w_{\text{tier}}$ of Section 4.1 (distinct from the CFG scale $w$ of Section 4.5).

Supplementary Fig. S14 walks the per-step mask construction. From cached eligibility $e \in \{0,1\}^4$ (renamed from $c$ to avoid collision with the conditioning vector) and the tier weight $w_{\text{tier}}$ of Section 4.1, five stochastic stages produce $m$: the *subset-size* box samples $k$ from a Categorical with probabilities $\{0{:}.10,\ 1{:}.25,\ 2{:}.30,\ 3{:}.25,\ 4{:}.10\}$; the *weighted pick* box samples $k$ properties from $e$ using $w_{\text{tier}}$-weighted multinomial draws to form a tentative one-hot; *property dropout* at rate 0.30 then independently zeros each entry of the tentative mask; *CFG dropout* at rate 0.10 zeros the entire mask to feed the unconditional branch.

The output $m$ and the row weight $\omega_{\text{row}} = \alpha + (1-\alpha) \cdot \mathrm{mean}(w_{\text{tier}} \odot m)$ are emitted to the denoiser-training step of Section 4.7, where $w_{\text{tier}}$ is the per-property tier-weight vector and $\alpha \in [0,1]$ is the trust-vs-uniform interpolant (default $\alpha = 1.0$, reducing to plain MSE; $\alpha < 1.0$ up-weights more-trusted rows). The mask $m$ gates only the FiLM property input; it does not appear in the loss.

### 4.7 Denoiser training

The denoiser learns the noise-predicting reverse process of a 1000-step variance-preserving DDPM in latent space, conditioned via FiLM on Section 4.6 mask + property vector. Classifier-free guidance dropout at training time enables the runtime trade-off between unconditional and target-property sampling controlled by the CFG scale w.

The denoiser $\epsilon_\theta(z_t, t, c, m)$ is a 44.6 M-parameter FiLM-modulated ResNet over $z_t \in \mathbb{R}^{1024}$ with 8 residual blocks of inner width 2048 (per-block: LayerNorm, Linear(1024 to 2048), FiLM($\gamma, \beta$ from cond), SiLU, Linear(2048 to 1024) with a residual add), a sinusoidal time embedding ($d_t = 256$), and a sinusoidal property-value embedding ($d_c = 64$) gated by the per-step mask $m \in \{0,1\}^4$ of Section 4.6 (full hyperparameters in Supplementary Table S3).

Supplementary Fig. S15 walks denoiser training. Per step: sample $t \sim \mathcal{U}\{1{:}T\}$ and $\varepsilon \sim \mathcal{N}(0, I)$; form $z_t = \sqrt{\bar{\alpha}_t}\, z_0 + \sqrt{1 - \bar{\alpha}_t}\, \varepsilon$ on the cosine $T = 1000$ DDPM schedule of Nichol & Dhariwal [17]; FiLM injects $(t,\, p \odot m)$; the network predicts $\hat{\varepsilon}$; the loss is the per-sample MSE $\|\varepsilon - \hat{\varepsilon}\|^2$ weighted by the row weight $\omega_{\text{row}}$ of Section 4.6. Optimiser AdamW, peak LR $10^{-4}$, cosine decay, batch 128, 20 epochs, exponential-moving-average (EMA) decay 0.999.

At sample time we use the standard guided estimate

$$\hat{\epsilon}_\theta(z_t, c) \;=\; \epsilon_\theta(z_t, t, \emptyset) \;+\; w\,(\epsilon_\theta(z_t, t, c) - \epsilon_\theta(z_t, t, \emptyset));$$

the production scale $w = 7$ is selected per Section 4.15 (Supplementary Fig. S22).

### 4.8 Two complementary denoisers

The headline pipeline targets $\rho$, $D$, $P$, and HOF simultaneously. Their labelled-corpus marginals have sharply different population statistics: the high-HOF tail and the high-$\rho DP$ tail barely overlap in latent space. A single denoiser must compromise between these tails. Either tail can be trained to saturate, but not both at once with one loss-weighting schedule. We therefore train two complementary denoisers, each tilted toward one tail via training-time oversampling, so that each saturates its own tail.

DGLD-H tilts toward the HOF tail (asymmetric high-tail oversampling on Tier-A/B, factor 5×, focused on heat of formation). DGLD-P tilts toward the $\rho DP$ tail (Tier-A/B-only conditioning, +5× high-tail oversampling on the top decile of the joint $\rho/D/P$ distribution, weighted_mask=true, property-dropout 0.30). Both are 44.6 M-parameter FiLM-ResNets with Section 4.7 architecture; only the training-data tilt differs. Hyperparameters of both checkpoints are listed in Supplementary Table S3.

Empirically, DGLD-H supplies high-HOF leads and DGLD-P supplies high-$\rho/P$ leads. Combination happens post-decode in Section 4.14 (Supplementary Fig. S21): we union the lanes' SMILES pools and run them once through the rerank funnel. The combination is a post-decode union of the two lanes' SMILES pools, with no weight interpolation or output averaging.

### 4.9 Labeling for guidance-head training

Each score-model head needs per-row training labels. We run six independent label-source pipelines once over the corpus, before any score-model training begins. Three feed active steering signals (Viability, Sensitivity, Hazard) with energetic-domain or rule-based labels; three feed auxiliary multi-task heads (Performance, SA, SC) with smoke-ensemble or drug-domain labels.

Each head plays a distinct role. *Viability* is the gatekeeper gradient that steers away from non-energetic regions; *Sensitivity* and *Hazard* are the safety axes the user cares about. These three are the active steering signals invoked at sample time. *Performance, SA*, and *SC* are auxiliary multi-task heads: trained on the shared trunk for regularisation but not invoked in the steering bus. Stage 2 of the Supplementary Fig. S20 reranker uses the Uni-Mol smoke ensemble (Section 4.13) for property scoring, not the latent Performance head.

Supplementary Fig. S16 walks the base label pipelines. A Random Forest on Morgan FP + RDKit descriptors yields $y_{\text{viab}}$; a Politzer–Murray BDE chemotype-class fit on Huang & Massa [24] $h_{50}$ data yields $y_{\text{sens}}$ (y_sens ∈ [0,1] with 1 = high sensitivity / dangerous; guidance descends this signal to reduce predicted sensitivity); the chemist-curated SMARTS catalog plus the Bruns–Watson demerit list [49] yields $y_{\text{haz}}$; and a Uni-Mol smoke ensemble yields $y_{\text{perf}} \in \mathbb{R}^4$ over $(\rho, D, P, \text{HOF})$. The cached LIMO $\mu$ is held separately for the noised-latent training of the score model. Random Forest probability is portable, smooth, and decoupled; see Supplementary Section B.1.

### 4.10 Score-model training

All six heads are trained jointly on a shared 4-block FiLM-MLP trunk under multi-task supervision. The trunk is regularised by all six losses; at sample time, only the three active steering signals are actually invoked. Auxiliary multi-task heads (Performance, SA, SC) earn their trunk-improving role during training but stay quiet during sampling.

The score-model trunk and six heads are trained on noised LIMO latents using the labels of Section 4.9 under multi-task supervision: three of the six (Viability, Sensitivity, Hazard) contribute steering signals at sample time, three (Performance, SA, SC) are auxiliary multi-task signals trained for trunk regularisation only.

Supplementary Fig. S17 walks score-model training. A shared 4-block FiLM-MLP trunk (1024-d hidden) consumes $(z_t, \sigma_t)$, where $\sigma_t = \sqrt{1 - \bar{\alpha}_t}$ is embedded into a 128-d sinusoidal token. Six heads branch off the trunk: Viability and Hazard are sigmoid heads trained with binary cross-entropy; Sensitivity, SA, and SC are smooth-L1 regressors; Performance is a 4-vector smooth-L1 regressor on $(\rho, D, P, \text{HOF})$. Crucially, training data are *noised* latents at uniform $t$, not clean $\mu$, so the gradient is queried at every $\sigma_t$ along the sampling trajectory. The total loss is the availability-mask-gated sum

$$\mathcal{L}_{\text{score}} = \sum_k a_k \cdot w_k \cdot \mathcal{L}_k(\hat{y}_k(z_t, \sigma_t),\ y_k),$$

where $a \in \{0,1\}^6$ is the per-head *availability* mask (distinct from the denoiser's per-property conditioning mask $m \in \{0,1\}^4$ of Supplementary Figs S14 and S15) and $w_k$ are static head weights chosen so each $\mathcal{L}_k$ sits at $\mathcal{O}(1)$ at convergence (values in Supplementary Table S3). Optimiser AdamW, peak LR $2 \times 10^{-4}$, cosine decay, batch 1024, ~40k steps, EMA decay 0.999.

The auxiliary multi-task heads (Performance, SA, SC) train on the shared trunk for regularisation only, not invoked in the production steering bus (Supplementary Section B.4). Two checkpoint variants (production 6-head hazard-aware and its 5-head predecessor) are released on Zenodo (Supplementary Section D.1).

### 4.11 Self-distillation refinement of the viability head

Round-0 viability labels combine SMARTS pass/fail with a Random Forest classifier trained on (energetic corpus, ZINC drug-like), a coarse two-class boundary. Self-distillation closes the gap between that boundary and the latent regions the diffusion sampler actually inhabits, by mining the model's own surrogate-gaming false positives and re-feeding them as labelled hard negatives.

The FROZEN banner of Supplementary Fig. S18 is read literally. Across all rounds the LIMO encoder, LIMO decoder, Section 4.8 denoisers, the Random Forest, and the SMARTS rulebook do not retrain. The score-model trunk and heads are the only thing that updates between rounds. Within the score model, hard-negative latents are labelled $y_{\text{viab}} = 0$ only; the other heads see those rows with $a_k = 0$ (unlabelled).

The five-step protocol (sample → mine → encode → retrain → probe) is detailed in Supplementary Section B.1. The hard-negative ticker on Supplementary Fig. S18 shows $0 \to 137 \to 918$ cumulative negatives over rounds 0/1/2. (round-0 training uses 0 hard negatives; round-0 mining produces 137; round-1 training uses 137; round-1 mining cumulates to 918; round-2 / production training uses 918.)

The stopping criterion is shown in the right-margin box of Supplementary Fig. S18: a held-out probe (7 anchors {RDX, HMX, TNT, FOX-7, PETN, TATB, NTO} and 5 gaming molecules) must show every anchor at ≥ 0.86 AND every gaming molecule at ≤ 0.84. Empirically 3 rounds satisfy this; round 2 (corpus + 918 hard negatives) is the production checkpoint. Pseudocode in Supplementary Section B.1.

### 4.12 Sampling

At sample time, three of the six trained heads (Viability, Sensitivity, Hazard) supply gradient steering signals at every DDIM (denoising diffusion implicit model) step on top of classifier-free guidance. The denoiser remains the same checkpoint as Section 4.7; only the sampler changes.

Supplementary Fig. S19 walks sampling. A latent $z_T \sim \mathcal{N}(0, I_{1024})$ is denoised over $t = T \to 1$ in 40 DDIM steps. At each step,

$$\hat{\epsilon} \; = \; \epsilon_\theta^{\text{cfg}}(z_t, t, c) \; - \; \sigma_t \sum_{h \in \{\text{viab,sens,hazard}\}} s_h \, \nabla_{z_t} \mathcal{L}_h(z_t, \sigma_t),$$

where $\epsilon_\theta^{\text{cfg}}$ is the standard CFG noise estimate over the frozen denoiser of Section 4.8 (Supplementary Fig. S15). The per-head losses are $-\log P_{\text{viab}}$ (ascend), $\text{sens}_z$ (descend), and $-\log(1 - P_{\text{hazard}})$ (descend hazard). Production scales are $s_{\text{viab}} = 1.0$, $s_{\text{sens}} = 0.3$, $s_{\text{hazard}} = 1.0$ (selection rationale in Section 4.15). The final $z_0$ decodes through the frozen LIMO decoder to a SMILES pool consumed by Supplementary Fig. S21.

The classifier-guidance steering bus uses two configurable knobs: a noise-dependent annealing factor $\alpha(\sigma_t) = \max(0,\, 1 - \sigma_t/\sigma_{\max})$ that scales the bus down at high noise levels (where the score-model heads have not yet seen enough latent structure to be reliable), and a per-row gradient-norm clamp at magnitude $C_g$ (gradient clamp) that prevents a single row from dominating the batch. The image-domain classifier-guidance recipe of Dhariwal & Nichol [17] uses $\sigma_{\max} = \sigma_T$ (full ramp on) and $C_g = 5.0$; this works well at 1000 sampling steps, where each step covers a small $\sigma$-jump and the natural per-row gradient magnitudes sit well below 5. In the 40-step latent regime the same recipe zeroes the bus at the top of the trajectory (the first ~10 steps live near $\sigma_t \approx \sigma_T$) and saturates the clamp at the bottom (natural gradient magnitudes at low $\sigma$ are 8–30, well above 5), so the per-head scales $s_h$ have negligible effect on the produced chemistry. Production disables the anneal by setting $\alpha \equiv 1$ directly (equivalent to the $\sigma_{\max} \to \infty$ limit of the schedule above; we configure this via a guard clause on $\sigma_{\max} = 0$ in code, treating it as the no-anneal sentinel) and loosens the clamp ($C_g = 50$). The four-test diagnostic that mapped per-step gradient norms across both configurations is in Supplementary Section D.5. The CFG scale $w$ is selected from Section 4.15 sweep (Supplementary Fig. S22, $w \in \{5,7,9\}$ at pool=8k); per-property quantile-error breakdown is in Supplementary Section D.8.

### 4.13 Filtering

The decoded SMILES pool flows through a four-stage funnel that progressively raises the bar from cheap chemistry rules (SMARTS gate) to expensive first-principles audits (DFT). Stages 1+2 (ms cost per candidate) score every candidate; Stages 3+4 (CPU- and GPU-hours per candidate) run only on the top-K survivors.

Supplementary Fig. S20 walks the four-stage filtering funnel. Each pool is canonicalised, deduplicated, and stripped of charged species and over-large molecules.

*Stage 1: SMARTS gate (rules + redflags).* A chemist-curated SMARTS [50] catalog removes radicals, sulphur, halogens, mixed valence states, and other red flags; survivors are scored by the Uni-Mol smoke ensemble. SA [35] ≤ 5.0 and SCScore [36] ≤ 3.5 caps are applied. Tanimoto similarity to the nearest training-set neighbour is required to lie within [0.20,0.55].

*Stage 2: Pareto reranker.* The composite

$$S(x) \;=\; 0.45\, S_{\text{perf}}^{\text{band}}(x) + 0.20\, S_{\text{viab}}(x) + 0.15\, S_{\text{novel}}(x) + 0.20\,(1 - S_{\text{sens}}(x)) - 0.10\, S_{\text{alerts}}(x)$$

is computed, then a Pareto front over (−perf, −viab, sens, alerts) is used as the outer stratifier with the composite as within-stratum tiebreak. Default top-$K = 100$. The Stage-2 reranker draws property scores ($\rho, D, P,$ HOF) from the Uni-Mol smoke ensemble (via evaluate_candidates.py + rerank_v2.py), *not* from the latent Performance head; the latent Performance head trains as a multitask auxiliary on the shared trunk for regularisation only and is not invoked at rerank time. *Stage 3: xTB triage.* GFN2-xTB optimisation, HOMO–LUMO gap ≥ 1.5 eV (Section 2.3). *Stage 4: DFT audit.* B3LYP/6-31G(d) optimisation + ωB97X-D3BJ/def2-TZVP single-point + 6-anchor calibration + Kamlet–Jacobs (Section 2.3). Illustrative survivors at the Section 2 production setting: pool=40k → ~1 800 chem-pass after Stage 1+2 (4.6 % Stage-1+2 pass rate; 966 survive the full Uni-Mol + composite filter, Supplementary Section F.2) → top-100 by composite re-rank → 11 DFT-validated lead cards.

The Pareto-front procedure in Stage 2 is the four-step sweep:

1. **Score** each candidate with the composite $S(x)$ defined above (perf-band, viability, novelty, sensitivity, alerts).
2. **Mark** each candidate as on-front or dominated, against the four-objective comparison ($-$perf, $-$viability, sensitivity, alerts) using a standard pairwise non-dominance sweep.
3. **Sort** with a two-key order: Pareto-front candidates first, then non-front; within each stratum by composite descending.
4. **Return** the top-$K$ (default $K = 100$).

### 4.14 Pool fusion across sampling runs

A single sampling lane fixes one (denoiser, conditions, guidance) tuple. Pool fusion runs multiple lanes in parallel and unions their decoded SMILES outputs, exploiting the lanes' independent failure modes for diversity. The Section 4.13 reranker is what tie-breaks across the fused pool.

Supplementary Fig. S21 walks pool fusion. A single sampling lane = one (denoiser, conditions, guidance) tuple, run end-to-end ($z_T$ draw → 40 DDIM with that lane's config → LIMO decode → SMILES pool). The production methodology recipe is two lanes, one per denoiser (DGLD-H and DGLD-P), both at the headline target conditions, both at viab+sens+hazard guidance, both at CFG $w = 7$, pool ≥ 40k each. Fusion is post-decode: the lanes' pools are unioned, canonical-SMILES deduplicated, and fed into the Stage-1 reranker. The three orthogonal diversity axes are conditions, denoisers, and guidance; each breaks a different correlated failure mode. The Supplementary Section F.4 four-pool merge (which adds two unguided ablation lanes) is a presentation choice for the merged top-100 result, not the production methodology recipe.

### 4.15 Training and hyperparameter selection

Every numeric constant in Section 4.5-Section 4.14 is selected by one of four mechanisms: empirically swept, stop-criterion-driven, inherited from prior work, or chemist-set. This section walks each bucket and flags what was empirically verified versus what was a defensible default.

Every numeric constant introduced in Section 4.5–Section 4.14 falls into one of four buckets, distinguished by *how* the value was set: empirical sweeps, stop-criterion-driven counts, values inherited from prior work, and chemist-set thresholds. The remainder of this section walks each bucket in turn, then closes with the limitations of the selection procedure.

The first bucket is empirically swept. Two knobs were swept directly under the post-filter survival metric: the classifier-free-guidance scale $w$ (Supplementary Fig. S22) and the candidate pool size (Supplementary Fig. S23). The CFG scale was tested at three points, $w \in \{5,7,9\}$ at pool=8k, ranked by post-filter yield; $w = 7$ is the empirical sweet spot, with more candidates surviving than at $w = 5$ and the tight-mode collapse at $w = 9$ avoided. The pool-size sweep ranged from 1.5k to 40k, with both the best composite and the post-filter survival count still climbing at 40k; production uses pool ≥ 40k per lane. The per-head guidance scales $s_h$ were also empirically chosen via the Supplementary Section F.4 multi-axis matrix (full grid in Supplementary Section D.6), and the score-head loss weights $w_k$ were hand-set so each $\mathcal{L}_k$ sits at $\mathcal{O}(1)$ at convergence (ablation in Supplementary Section B.3).

The second bucket is stop-criterion-driven. The self-distillation round count was set by the held-out probe described in Section 4.11, which requires every anchor at $\geq 0.86$ and every gaming molecule at $\leq 0.84$; the production budget-918 checkpoint is round 2 of self-distillation (round 0 = score-model trained on corpus only with Random-Forest-derived viability labels and 0 hard negatives, round 1 = corpus + 137 mined hard negatives, round 2 = corpus + 918 cumulative hard negatives + aromatic-heterocycle boost), the configuration that demotes the worst gaming molecule (gem-tetranitro) furthest, to viability 0.74. The hard-negative count (918) is the cumulative round-2 mining yield, not a tuned target.

The third bucket is inherited from prior work or community convention and is not retuned in this paper. The KL weight $\beta = 0.01$ is the LIMO original; the DDPM uses the cosine $T = 1000$ schedule of Ho et al. [14]; the per-property dropout rate of 0.30 follows FiLM convention; the CFG dropout rate of 0.10 follows Ho and Salimans 2022; AdamW with peak LR $10^{-4}$ on a cosine schedule plus EMA decay 0.999 is the standard diffusion-training recipe.

The fourth bucket is chemist-set thresholds, fixed by domain conventions rather than by sweep. The xTB HOMO–LUMO cut is $\geq$ 1.5 eV; the DFT calibration set is the 6-anchor panel (RDX, TATB, HMX, PETN, FOX-7, NTO); the Tanimoto novelty window is $[0.20, 0.55]$ (operational novelty criterion for this pipeline; absolute scaffold novelty in the high-energy-density-materials (HEDM) literature requires additional expert review); molecular-weight floor is 130 Da and the oxygen-balance cap is $+25$ %.

The selection procedure uses 1D ablations throughout: each hyperparameter is fixed by a per-axis sweep, a stop-criterion count, a prior-art default, or a chemist-set threshold, with the full grids in Section 2 and Supplementary Section D. A joint optimisation over ($w$, $s_h$, pool size) and per-target retuning are scoped as future work. Section 2.6 hosts only the *use* of these settings (baseline comparisons, top-1 metrics) and Supplementary Section D gives the full per-axis grids; the selection process lives here. The full compute footprint is summarised in Supplementary Section D.2.

### 4.16 Baseline generators

Four no-diffusion baselines are run through the identical downstream validation pipeline (canonicalisation, chemistry filters, Uni-Mol surrogate scoring, novelty window) on the same training corpus, so that any difference reflects the generator rather than the evaluation.

**SMILES-LSTM.** A character-level LSTM (embedding → 2-layer LSTM, 512 hidden units, ~6 M parameters) trained from scratch on the 326 k-molecule energetic SMILES corpus (AdamW, learning rate $10^{-3}$, weight decay $10^{-4}$, batch 128, gradient clip 1.0, 5 epochs, sequence length $\leq$ 120). Ten thousand SMILES are sampled autoregressively at temperature 1.0 for each of three seeds. This baseline matches DGLD's corpus and parameter budget but omits diffusion and score-model guidance.

**SELFIES-GA.** A genetic algorithm over SELFIES (selfies 2.1.1: token substitution/insertion/deletion up to three mutations, single-point crossover), initialised from the same corpus and optimising the DGLD Pareto composite (viability, sensitivity, novelty, performance) each generation. The primary run evolves a population of 2 000 over 30 generations (elite fraction 0.50, new fraction 0.20, crossover fraction 0.15, seed 42); a compute-matched variant evolves 40 000 over 15 generations. Every candidate is scored each generation by the same Uni-Mol surrogate; the 40 000-pool best-novel candidate is carried through the full B3LYP/6-31G(d) → $\omega$B97X-D3BJ/def2-TZVP → Kamlet–Jacobs DFT audit.

**MolMIM 70 M.** NVIDIA MolMIM (~70 M parameters, molmim_70m_24_3 checkpoint, BioNeMo 1.5), used pretrained without energetic-domain fine-tuning as an out-of-domain reference. Molecules are produced by greedy–perturbate latent decoding (scaled radius 1.0, single seed), yielding 5 598 samples that are scored on the same surrogate. Its composite is on an uncalibrated scale and is not comparable to the other methods.

**REINVENT 4.** AstraZeneca REINVENT 4 (v4.4.12) with reinforcement-learning fine-tuning of its built-in SMILES-transformer prior under an N-fraction-oriented reward (N-rich substructure match, synthetic-accessibility penalty, molecular-weight floor, CHNO alerts), with no DGLD-corpus pretraining. The 40 000-target run uses 8 000 reinforcement-learning steps; a multi-seed variant uses staged learning (2 000 steps, DAP, $\sigma = 128$, seeds 42/1/2). Because the reward does not target $\rho/D/P$ directly, the top-100 by N-fraction are scored post hoc on the same surrogate.

All four baselines are scored on the same Uni-Mol surrogate; only the SELFIES-GA best-novel candidate is carried to DFT. The reranker composite is generator-internal for SELFIES-GA, an N-fraction proxy for REINVENT 4, and uncalibrated for MolMIM, so composite values are compared within, not across, methods.

### 4.17 Use of large language models

A large language model (Anthropic Claude) was used to assist with source-code development and manuscript copy-editing. All study design, analyses, results, and conclusions are the authors' own, and the authors take full responsibility for the content of the manuscript.

### Data Availability Statement

All code, data, and trained models are released together as a single archived package on Zenodo at https://zenodo.org/records/21735142 (code under Apache-2.0; data and model checkpoints under CC-BY-4.0). The package contains: the code (LIMO fine-tuning, denoiser and score-model training, sampling, the four-stage validation funnel, and the figure-generation pipeline); the trained checkpoints (the LIMO VAE, two conditional latent denoisers DGLD-H and DGLD-P, two multi-head latent score models, the SELFIES alphabet, and run metadata); and the data (the 65,980-row labelled CHNO master, the augmented unlabelled corpus, and the 918 mined hard-negative latents), redistributed in canonicalised form with row-level provenance.